\documentclass[prd,aps,preprintnumbers,showpacs,superscriptaddress]{revtex4}

\usepackage{amsmath}
\usepackage{array}
\usepackage{amssymb}
\usepackage{color}
\usepackage{epsfig}
\usepackage{graphicx}
\usepackage{psfrag}
\usepackage{slashed}

\newcommand{\be}{\begin{equation}}
\newcommand{\ee}{\end{equation}}
\newcommand{\bea}{\begin{eqnarray}}
\newcommand{\eea}{\end{eqnarray}}

\newcommand{\bm}[1]{\mbox{\boldmath $#1$}}

\newcommand{\st}{{\scriptscriptstyle T}}
\newcommand{\sL}{{\scriptscriptstyle L}}
\DeclareMathOperator{\tr}{Tr}

\begin{document}

\preprint{NIKHEF 2013-019}
\title{Generalized Universality of Definite Rank Gluon Transverse Momentum Dependent Correlators}

\author{M.G.A.~Buffing}
\email{m.g.a.buffing@vu.nl}
\affiliation{Nikhef and Department of Physics and Astronomy, VU University Amsterdam,\\
De Boelelaan 1081, NL-1081 HV Amsterdam, the Netherlands}

\author{A.~Mukherjee}
\email{asmita@phy.iitb.ac.in}
\affiliation{Department of Physics, Indian Institute of Technology Bombay, Powai, Mumbai 400076, India}
\affiliation{Institut f\"ur Theoretische Physik, Universit\"at T\"ubingen, 72076 T\"ubingen, Germany}

\author{P.J.~Mulders}
\email{mulders@few.vu.nl}
\affiliation{Nikhef and Department of Physics and Astronomy, VU University Amsterdam,\\
De Boelelaan 1081, NL-1081 HV Amsterdam, the Netherlands}

\begin{abstract}
Transverse momentum dependent (TMD) parton correlators describing the partonic structure of hadrons contain gauge links, required by color gauge invariance. The required gauge links enter in the matrix elements that contain the parton fields and depend on the color flow in the hard process. The correlators are expanded in terms of transverse momentum dependent parton distribution functions, referred to as TMD PDFs, or in short TMDs. In this paper, we introduce gluon TMDs of definite rank, by making an expansion of the TMD gluon correlator with the help of irreducible tensors built from the transverse momenta. The process dependence is isolated in gauge link dependent gluonic pole factors multiplying the TMDs. It is important to account for the different possibilities in the color structure within the matrix elements, leading to multiple TMDs at a given rank. In this way we are able to write the leading tree level result for a hard process in terms of process dependent gluon correlators which are expressed in a finite set of universal TMDs. We tabulate the gluonic pole factors for various gauge links, among them those that are relevant for $2\rightarrow 2$ processes.
\end{abstract}
\date{\today}

\maketitle

\section{Introduction}
The cross sections and spin asymmetries of hadronic scattering processes can be described in terms of `soft' functions like the distribution and fragmentation functions for quarks and gluons, convoluted with cross sections for `hard' partonic subprocesses. To leading order in an appropriate expansion involving the inverse hard scale relevant in the process, this description corresponds to the tree level result of the full treatment~\cite{Collins:2011zzd}. The distribution functions have the interpretation as the probability of finding a quark or a gluon with momentum fraction $x$ inside a hadron and fragmentation functions are interpretated as decay functions giving the number of hadrons with momentum fraction $z$ in the partonic decay chain; both appear in the parametrization of the parton correlators, which on the other hand are nonlocal matrix elements of partonic field operators. Together with (leading) collinear gluons exchanged between the soft and hard parts of the process, these correlators are gauge invariant with the collinear gluons showing up as Wilson lines or gauge links. Going beyond the collinear limit, one includes the dependence on the relative transverse momenta of the partons and hadrons in the correlators, which are then parametrized in terms of transverse momentum dependent (TMD) distribution and fragmentation functions. The TMD correlators contain operators that are nonlocal not only in the light-cone direction but also in the transverse direction and there is no unique way to connect the fields through a gauge link~\cite{Collins:2002kn,Belitsky:2002sm,Boer:2003cm}. This gauge link becomes dependent on the process under consideration. As has been shown in the Refs.~\cite{Brodsky:2002cx,Brodsky:2002rv,Bacchetta:2005rm}, the structure of the gauge link, including transverse pieces at light-cone infinity, plays a very important role in the understanding of single spin asymmetries at high energies. However, the presence of the gauge link complicates the study of factorization in processes involving TMD correlators~\cite{Collins:2008sg,Rogers:2010dm,Collins:2011ca,Buffing:2011mj,Rogers:2013zha} and it is important to understand the universality issues of them. Towards this goal, we note that the TMD correlators can be ordered according to their $p_\st$-dependence, which can conveniently be written in terms of symmetric traceless tensors constructed from $p_\st$. We use this to construct TMD correlators of a particular rank. For a given rank, one can construct collinear correlators only depending on a momentum fraction $x$ by weighting the TMDs with the appropriate tensor. We refer to the $p_\st$-moments as (collinear) transverse moments.

The rank 0 correlators contain the $x$ and $p_\st^2$-dependent functions that survive after integrating over azimuthal angles and rank 1 correlators are important for azimuthal asymmetries of the form $\sin (\varphi)$ or $\cos (\varphi)$. They can be distinguished in time-reversal even (T-even) and time-reversal odd (T-odd) correlators. The transverse moments of the T-odd correlators contain a quark-quark-gluon operator combination with vanishing gluon momentum referred to as gluonic pole or Efremov-Teryaev-Qiu-Sterman (ETQS) matrix elements~\cite{Efremov:1981sh,Efremov:1984ip,Qiu:1991pp,Qiu:1991wg,Qiu:1998ia,Kanazawa:2000hz}. Correlators containing gluonic poles appear in cross sections with multiplicative gluonic pole factors depending on the hard part of the process under consideration. It has been shown that for fragmentation the gluonic pole matrix elements vanish~\cite{Gamberg:2010gp,Meissner:2008yf,Gamberg:2008yt,Collins:2004nx,Metz:2002iz} and the T-odd effect arises purely from the fact that one is dealing with non-plane wave final states.

Higher rank correlators and transverse moments are important for the azimuthal asymmetries of the form $\cos (m\varphi)$ and $\sin (m\varphi)$. In a recent publication, we analyzed the rank two correlators and double transverse moments of quark TMD correlators and their parametrization in terms of quark TMD distribution and fragmentation functions depending on $x$ and $p_T^2$ for unpolarized and polarized spin 1/2 and spin 1 hadrons~\cite{Buffing:2012sz}. All correlators containing gluonic pole matrix elements appear with (calculable) gluonic pole factors depending on the color flow structure in the hard process, including at rank 2 also T-even (double gluonic pole) correlators.

In this paper, we extend our analysis to gluon TMD correlators, which gives rise to a richer color structure phenomenology due to the more complicated gauge link structures involved. By using similar formalisms that we used before for the classification of quark TMDs, we will systematically classify all allowed color structures and indicate how they give rise to a finite number of universal TMD correlators. In any hard scattering process, the TMDs involved can then be written as a process dependent combination of these universal TMDs, multiplied with gluonic pole coefficients. We will tabulate the gluonic pole factors for a representative set of gauge links, among them those that are relevant for $2\rightarrow 2$ processes. Finally we will tabulate the (gauge invariant) definitions of the universal correlators needed for the description of TMDs.

\section{Gauge links and color structures}\label{s:gaugelinks}
For TMD gluon distribution (and fragmentation) functions, the gauge link dependence is a crucial part. Gauge link dependent gluon correlators have been introduced before~\cite{Mulders:2000sh,Bomhof:2006dp,Bomhof:2007xt}
\be
\Gamma^{[U,U^\prime]\,\mu\nu}(x,p_\st;n) ={\int}\frac{d\,\xi{\cdot}P\,d^2\xi_\st}{(2\pi)^3}\ e^{ip\cdot\xi}
\,\langle P{,}S|\,F^{n\mu}(0)\,U_{[0,\xi]}^{\phantom{\prime}}\,F^{n\nu}(\xi)\,U_{[\xi,0]}^\prime\,|P{,}S\rangle\biggr|_{\text{LF}} ,
\ee
where the fields are matrix-valued and one still needs to make the result a color singlet by appropriate color tracing. The gauge link structure is denoted as superscript $[U,U^\prime]$ in the correlator $\Gamma^{[U,U^\prime]}$, also simply denoted as $\Gamma^{[U]}$. The relevant gauge link structure arises from a resummation of leading Feynman diagrams including collinear $n{\cdot}A$ gluons and they are built from staple-like links in combination with loops that will be color traced in various ways. Staple-like links are the Wilson lines $U_{[0,\xi]}^{[\pm]} = U_{[0,\pm\infty]}^{[n]} U_{[0_\st,\xi_\st]}^{\st}U_{[\pm \infty,\xi]}^{[n]}$ running from $0$ to $\xi$, in the arguments of $\Gamma$ simply referred to as $\pm$, and their Hermitian conjugates, $\pm^\dagger$, running from $\xi$ to $0$ as discussed in Ref.~\cite{Bomhof:2004aw}. Loops like $U^{[\square]}=U_{[0,\xi]}^{[+]}U_{[\xi,0]}^{[-]}$ = $U_{[0,\xi]}^{[+]}U_{[0,\xi]}^{[-]\dagger}$ or $U^{[\square]\dagger}$ = $U_{[0,\xi]}^{[-]}U_{[\xi,0]}^{[+]}$ = $U_{[0,\xi]}^{[-]}U_{[0,\xi]}^{[+]\dagger}$ are referred to as $\square$ or as $\square^\dagger$. Besides a single trace, there are other color combinations that appear. The gluon correlator is bilocal with $F^{n\mu}(0)\,F^{n\nu}(\xi)$ containing the basic nonlocality $\xi$ that transforms into the parton (gluon) momentum $p$. Connecting the two nonlocal gluon fields, one has different types of gauge link structures,
\bea
\text{type 1:}&&\hspace{5mm}\tr_c \Big\lgroup F^{n\mu}(0)\,U_{[0,\xi]}^{\phantom{\prime}}\,F^{n\nu}(\xi)\,U_{[\xi,0]}^\prime\Big\rgroup, \label{e:type1} \\
\text{type 2:}&&\hspace{5mm}\tr_c \Big\lgroup F^{n\mu}(0)\,U_{[0,\xi]}^{\phantom{\prime}}\,F^{n\nu}(\xi)\,U_{[\xi,0]}^\prime\Big\rgroup\,\frac{1}{N_c}\tr_c \Big\lgroup U^{[\text{loop}]}\Big\rgroup, \label{e:type2} \\
\text{type 3:}&&\hspace{5mm}\frac{1}{N_c}\tr_c \Big\lgroup F^{n\mu}(0)\,U^{[\text{loop}]}\Big\rgroup\,\tr_c \Big\lgroup F^{n\nu}(\xi)\,U^{[\text{loop}^{\prime}]}\Big\rgroup. \label{e:type3}
\eea
These three types of contributions will be discussed below. Without the parton-like gluon fields defining the nonlocality, we could in principle even add a structure of the form
\bea
\text{type 0:}&&\hspace{5mm}\frac{1}{N_c}\tr_c \Big\lgroup U^{[\text{loop}]}
\Big\rgroup , 
\label{e:type0}
\eea
which could play a role in diffractive scattering or saturation~\cite{Dominguez:2010xd,Dominguez:2011wm}. We will restrict ourselves in this paper to the two-gluon types.

\subsection{Correlators of the first type}
\begin{figure}
\epsfig{file=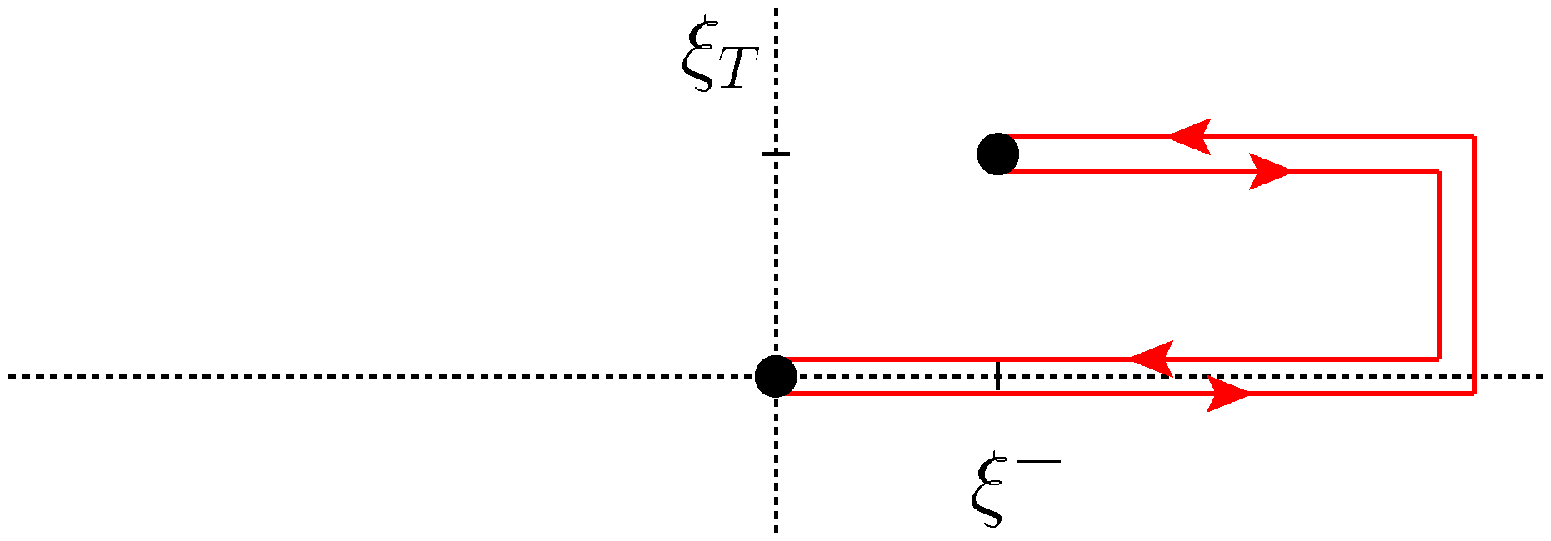,width=0.35\textwidth}
\hspace{15mm}
\epsfig{file=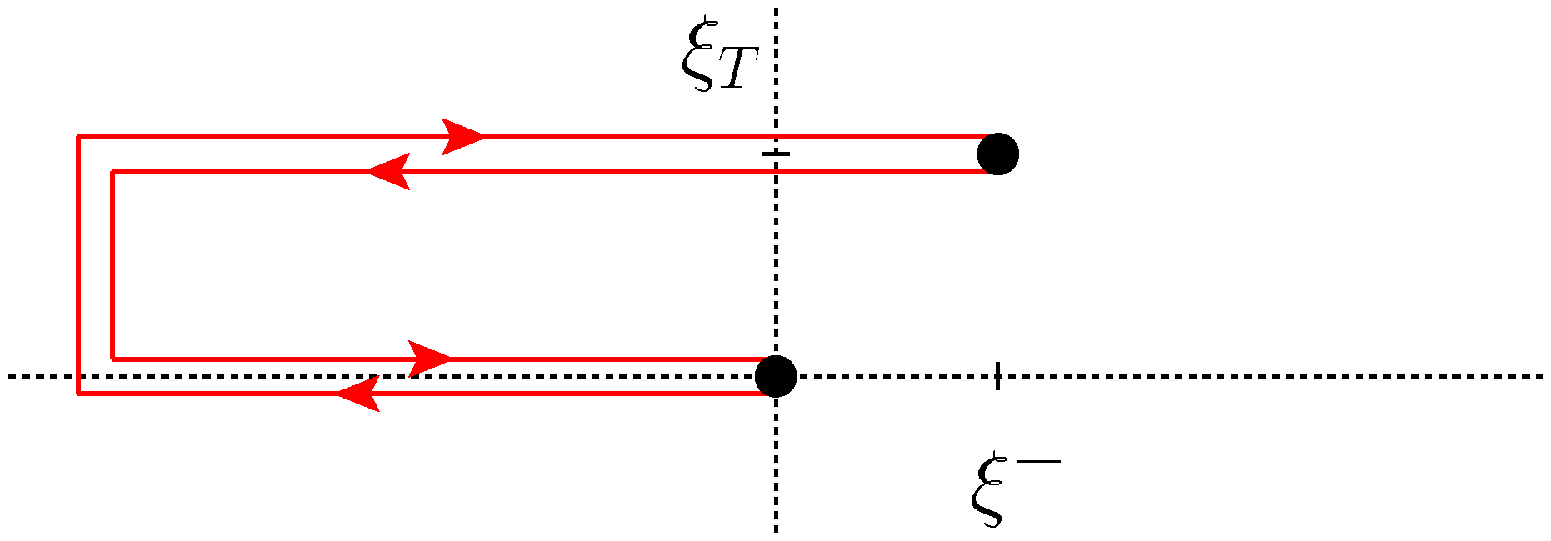,width=0.35\textwidth}
\\[1mm]
(a)\hspace{77mm} (b)
\\[3mm]
\epsfig{file=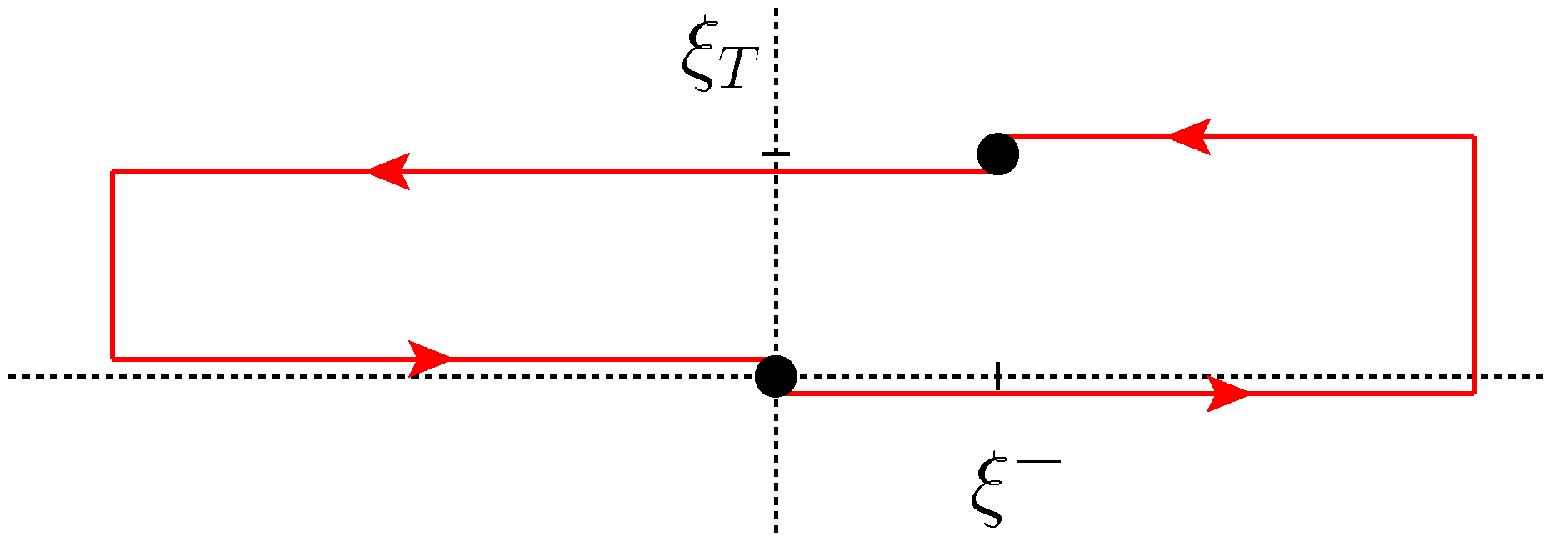,width=0.35\textwidth}
\hspace{15mm}
\epsfig{file=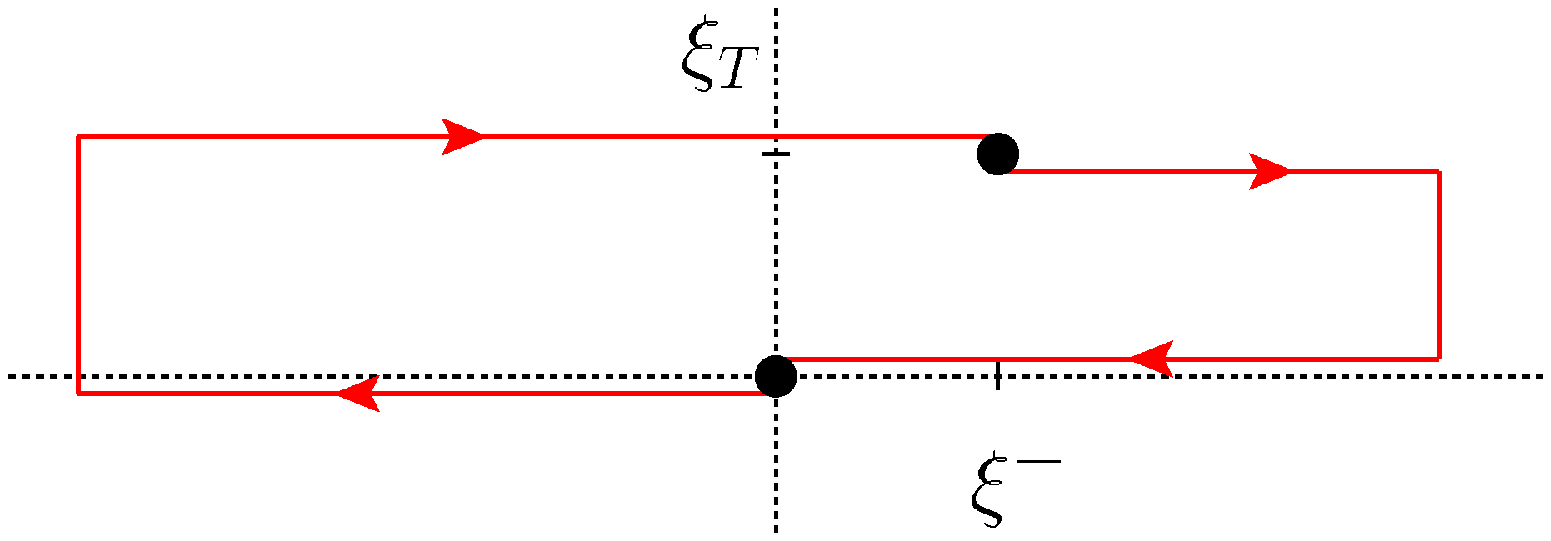,width=0.35\textwidth}
\\[1mm]
(c)\hspace{77mm} (d)
\\[3mm]
\epsfig{file=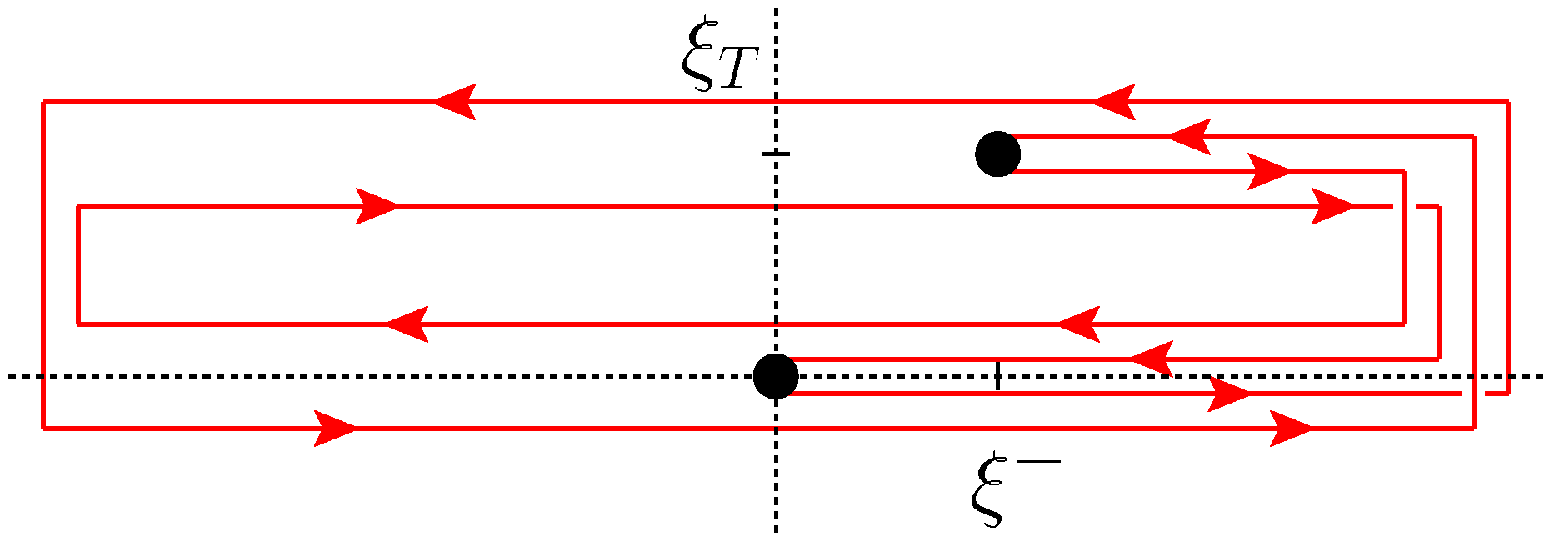,width=0.35\textwidth}
\hspace{15mm}
\epsfig{file=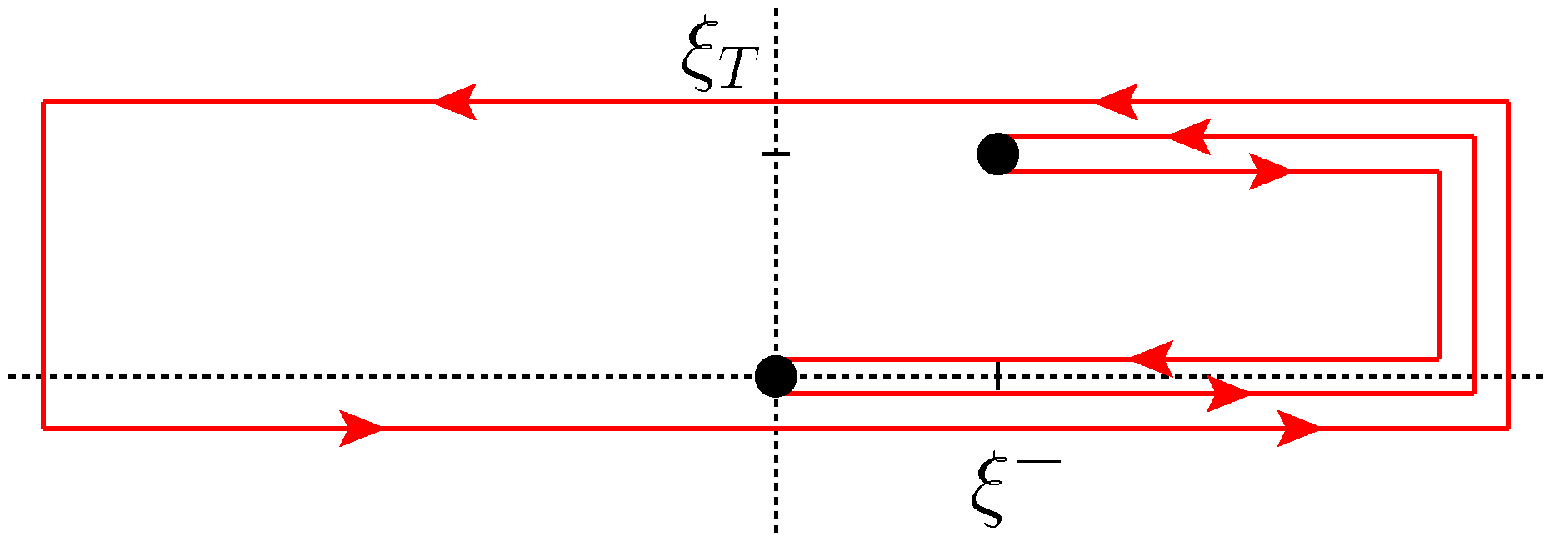,width=0.35\textwidth}
\\[1mm]
(e)\hspace{77mm} (f)
\\[3mm]
\epsfig{file=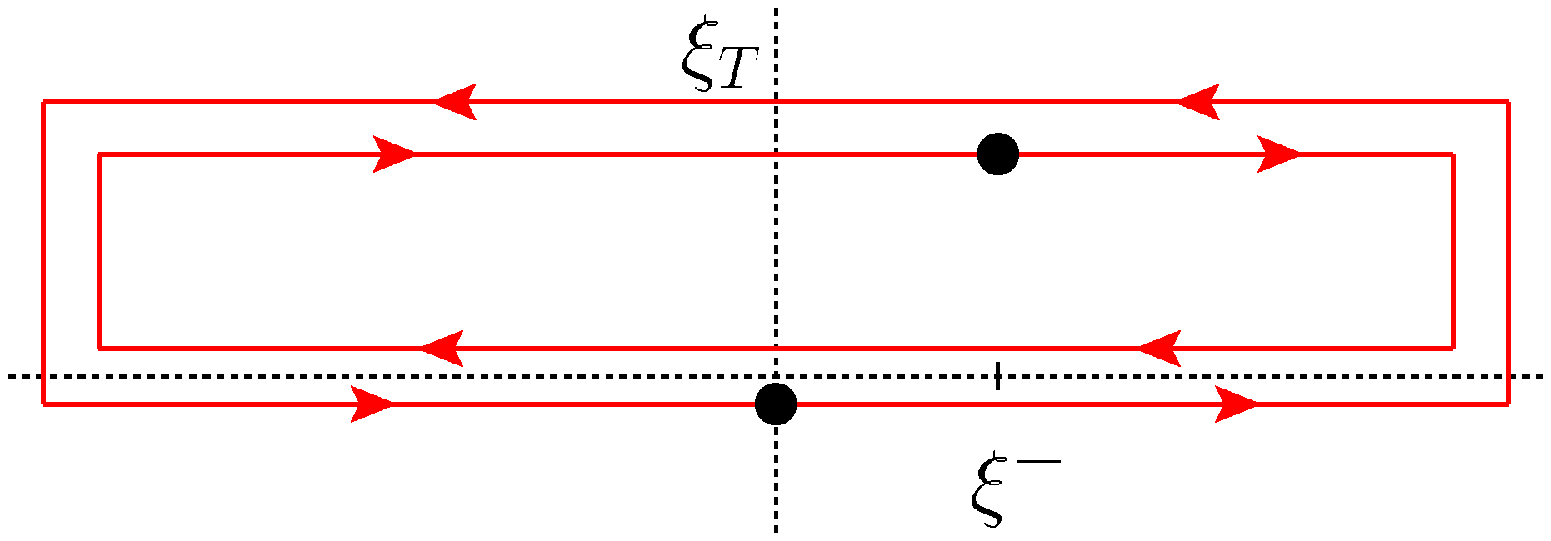,width=0.35\textwidth}
\hspace{15mm}
\epsfig{file=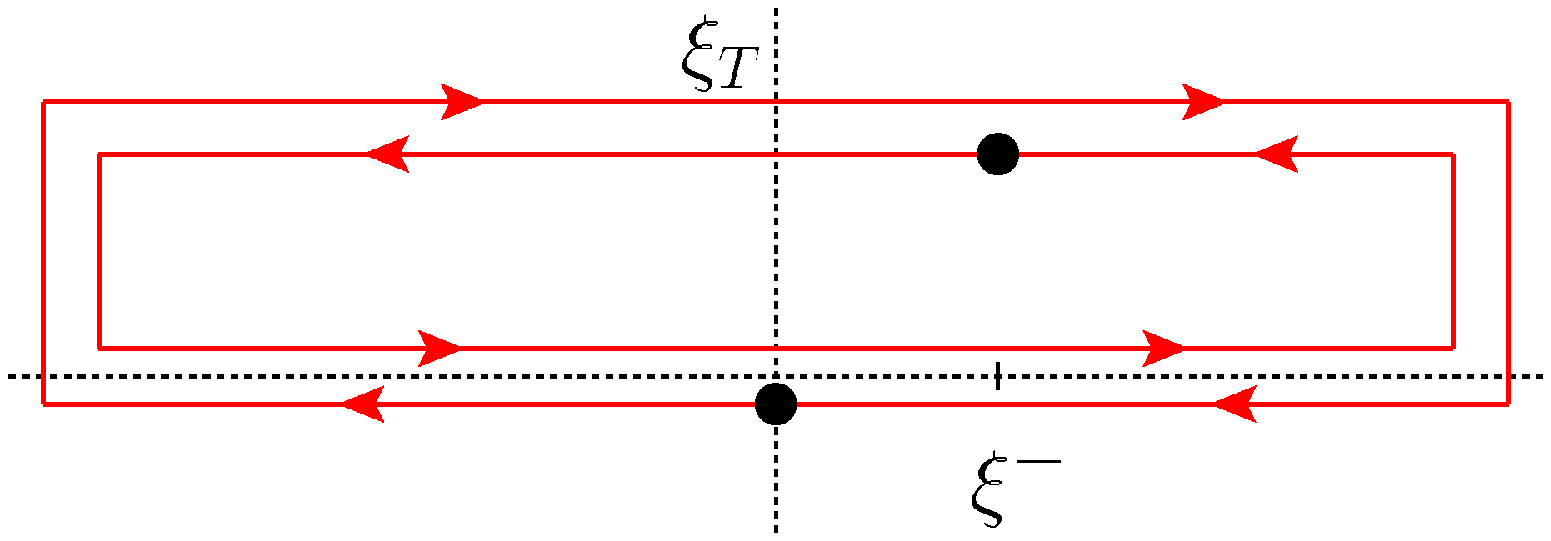,width=0.35\textwidth}
\\[1mm]
(g)\hspace{77mm} (h)
\caption{\label{f:GL}
A number of gauge link structures $[U,U^\prime]$ illustrated. In these figures, the two big dots represent the coordinates of $0$ and $\xi$. The horizontal axis is the light-cone direction $n_-$ and the vertical axis represents the transverse directions. The four simplest gauge link structures are 
(a) the $[+,+^{\dagger}]$ gauge link, 
(b) the $[-,-^{\dagger}]$ gauge link, 
(c) the $[+,-^{\dagger}]$ gauge link and 
(d) the $[-,+^{\dagger}]$ gauge link. 
Another structure occurring for type 1 correlators is the 
(e) $[+\square,+^{\dagger}\square^{\dagger}]$ gauge link. In 
(f) the type 2 gauge link structure $[+,+^{\dagger}(\square)]$ 
can be seen. Two gauge link structures corresponding to correlators 
of the third type are 
(g) $[(F(0)\square),(F(\xi)\square^{\dagger})]$ and 
(h) $[(F(0)\square^{\dagger}),(F(\xi)\square)]$.}
\end{figure}

The first type of operator structures contains one color trace and therefore has the simplest gauge link structure that is possible for gluon correlators. This color trace contains both the gluon field operators $F(0)$ and $F(\xi)$ with gauge links running between these two fields and appears when the color in a diagram contributing to the full amplitude flows in just a single color loop. Examples of processes containing such a color structure are processes with colorless particles in the final state, such as the gluon Drell-Yan process and Higgs production through gluon fusion ($gg\rightarrow H$), of which the color flow is simple (we will come back to this later), resulting in the gauge link structure in Fig.~\ref{f:GL}b. For these processes the gauge links run through minus light-cone infinity and the notation $\Gamma^{[-,-^{\dagger}]}$ is used to describe the link dependence of the correlator. In other processes, one or both of the gauge links could run through plus light-cone infinity, giving the additional gauge link structures $\Gamma^{[+,-^{\dagger}]}$, $\Gamma^{[-,+^{\dagger}]}$ and $\Gamma^{[+,+^{\dagger}]}$. The latter occurs for example when all color flows into the final state, e.g. in photon-gluon fusion producing quark-antiquark pairs, in which case both gauge links run through plus light-cone infinity, involving the correlator $\Gamma^{[+,+^{\dagger}]}$. Another relevant gauge link structure of the first type that occurs in a leading order $2\rightarrow 2$ process is the gauge link structure in the correlator $\Gamma^{[+\square,+^{\dagger}\square^{\dagger}]}$, which is illustrated in Fig.~\ref{f:GL}e.

\subsection{Correlators of the second type\label{s:cortype2}}
The second type of operator structures contains correlators with multiple color loops, where the two gluon fields are located in the same color trace. Typical diagrams where multiple color loops appear are diagrams with colored particles originating from two initial state hadrons and two colored particles being part of the final state. For an individual diagram, usually multiple color flow possibilities have to be taken into account for a correct description of the process. For this reason multiple different gauge link structures appear for most of the diagrams in $2\rightarrow 2$ processes, as can be seen explicitly in the appendix of Ref.~\cite{Bomhof:2006dp}.

In the correlator, the color traces that are without gluon fields all contain a gauge link loop (also called Wilson loop) $U^{[\square]}$ or $U^{[\square]\dagger}$. In principle, even more traced Wilson loops could appear, provided that the diagram has enough complexity to allow for multiple color loops to be present. The most complicated diagram that is relevant for our purposes contains three color loops. Since the color trace of a Wilson loop is a color singlet, it is possible to move the color traced Wilson loops to other places within the matrix element of the correlator. Gauge link structures of this second type that are needed in physical processes are $\Gamma^{[+(\square),+^{\dagger}]}$ = $\Gamma^{[+,+^{\dagger}(\square)]}$, $\Gamma^{[+(\square^{\dagger}),+^{\dagger}]}$ = $\Gamma^{[+,+^{\dagger}(\square^\dagger)]}$, $\Gamma^{[+,+^\dagger (\square)(\square^\dagger)]}$. In Fig.~\ref{f:GL}f the gauge link structure for $\Gamma^{[+,+^{\dagger}(\square)]}$ is illustrated. It should be noted that not all of the above mentioned structures occur if Wilson lines running through minus light-cone infinity instead of plus light-cone infinity are considered, as was the case for the four simplest type 1 gauge links. The only additional link structures in gluon correlators for leading $2\rightarrow 2$ processes are $\Gamma^{[+,-^{\dagger}(\square^\dagger)]}$ and $\Gamma^{[-,+^{\dagger}(\square)]}$.

\subsection{Correlators of the third type}
The third type of correlators has two color traces. Each color trace contains one of the two gluon fields and a Wilson loop, indicated by the $U^{[\text{loop}]}$ and $U^{[\text{loop}^{\prime}]}$ in Eq.~\ref{e:type3}. There are two relevant combinations, namely $U^{[\text{loop}]}=U^{[\square]}$ and $U^{[\text{loop}^{\prime}]}=U^{[\square^\dagger]}$, or the interchange of these. In the Figs.~\ref{f:GL}g and~\ref{f:GL}h these two structures are illustrated. Correlators that contain this gauge link structure correspond to a color flow where the (hadron) correlators are color singlet at the cut. In principle more complicated gauge link structures could be written down, but only if diagrams are included which allow for more complicated color flow structures, which do not occur for leading order $2\rightarrow 2$ diagrams. Diagrams containing three or more particles in the final state are less relevant for our study, because they usually don't allow measurements of (small) transverse momentum components.

\section{Formalism}\label{s:formalism}
In this section, we will give the gluon TMD distribution functions.
After this, we will give the procedure for taking transverse moments of gluon correlators. In this procedure, the building blocks of the matrix elements will be introduced naturally.

\subsection{Parametrization of gluon distribution functions}
In the previous section, the correlator has been defined in terms of matrix elements, which cannot be calculated from first principles. The correlator can also be written down by writing an expansion in TMD PDFs, first given in Ref.~\cite{Mulders:2000sh}, which following the naming convention in Ref.~\cite{Meissner:2007rx} is given by
\bea
2x\,\Gamma^{\mu\nu [U]}(x{,}p_\st) &=& 
-g_T^{\mu\nu}\,f_1^{g [U]}(x{,}p_\st^2)
+g_T^{\mu\nu}\frac{\epsilon_T^{p_TS_T}}{M}\,f_{1T}^{\perp g[U]}(x{,}p_\st^2)
\nonumber\\&&
+i\epsilon_T^{\mu\nu}\;g_{1s}^{g [U]}(x{,}p_\st)
+\bigg(\frac{p_T^\mu p_T^\nu}{M^2}\,{-}\,g_T^{\mu\nu}\frac{p_\st^2}{2M^2}\bigg)\;h_1^{\perp g [U]}(x{,}p_\st^2)
\nonumber\\ &&
-\frac{\epsilon_T^{p_T\{\mu}p_T^{\nu\}}}{2M^2}\;h_{1s}^{\perp g [U]}(x{,}p_\st)
-\frac{\epsilon_T^{p_T\{\mu}S_T^{\nu\}}{+}\epsilon_T^{S_T\{\mu}p_T^{\nu\}}}{4M}\;h_{1T}^{g[U]}(x{,}p_\st^2).
\label{e:GluonCorr}
\eea
In this parametrization, the spin vector is parametrized as $S^\mu = S_{\sL}P^\mu + S^\mu_{\st} + M^2\,S_{\sL}n^\mu$ and we used shorthand notations $g_{1s}^{g [U]}$ and $h_{1s}^{\perp g [U]}$,
\be
g_{1s}^{g [U]}(x,p_\st)=S_{\sL} g_{1L}^{g [U]}(x,p_{\st}^2)-\frac{p_{\st}\cdot S_{\st}}{M}g_{1T}^{g [U]}(x,p_{\st}^2).
\ee
Also the tensor $\epsilon_T^{\mu\nu}=\epsilon^{nP\mu\nu} = \epsilon^{\rho\sigma\mu\nu}n_{\rho}P_{\sigma}$ is used, where we have used shorthand notations like $\epsilon_T^{p_T \nu}=\epsilon_T^{\mu\nu}p_{T\mu}$. All possible types of two-gluon correlators mentioned before can be parametrized in this way. The TMDs $f_{1T}^{\perp g}$, $h_{1T}^{g}$, $h_{1L}^{\perp g}$ and $h_{1T}^{\perp g}$ are {\em naive} T-odd. This implies that they have the behavior $f_{1T}^{\perp g[U]}=-f_{1T}^{\perp g[U^t]}$, where $U^t$ is a time-reversed gauge link, interchanging plus and minus light-cone infinity~\cite{Boer:2003cm}. In the $p_\st$-integrated version of Eq.~\ref{e:GluonCorr} only $f_1^g$ and $g_{1L}^g$ survive, giving the well-known collinear gluon PDFs $g(x) = f_1^g(x)$ and $\Delta g(x) = g_{1L}^g(x)$. These collinear functions are universal as all gauge links reduce to a unique link along $n$. The TMDs that do not survive the $p_\st$-integration in Eq.~\ref{e:GluonCorr} all are multiplied by tensors containing $p_\st$ in symmetric traceless combinations. In order to reduce these terms in the correlator to a collinear form, one has to perform specific transverse weightings multiplying the correlator with additional factors of $p_\st$. Just as found for quark correlators, one then obtains expressions containing the TMD PDFs in Eq.~\ref{e:GluonCorr} weighted with powers of $-p_\st^2/2M^2 = \bm p_\st^2/2M^2$, denoted
\be
f_{\ldots}^{g (m)}(x,p_\st^2) = \left(\frac{-p_\st^2}{2M^2}\right)^m\,f_{\ldots}^{g}(x,p_\st^2).
\label{e:transversemoments}
\ee
For the explicit calculation tensor product relations are used. Depending on the rank of the tensor structure in Eq.~\ref{e:GluonCorr} one needs a corresponding number of transverse momenta $p_\st$ multiplying the correlator $\Gamma (x,p_\st)$. With single transverse weighting and azimuthal averaging of the correlator one obtains $f_{1T}^{\perp g (1)}$, $g_{1T}^{g (1)}$ and $h_{1T}^{g (1)}$. One obtains $h_1^{\perp g (2)}$ and $h_{1L}^{\perp g (2)}$ with double transverse weighting and azimuthal averaging of the correlator. Finally, a triple transverse weighting and azimuthal averaging of the correlator is required to obtain $h_{1T}^{\perp g (3)}$.

In the next subsection, we will show how the transverse weightings are performed at the level of the matrix elements including in particular the role of the gauge link structure in this. The TMDs in Eq.~\ref{e:GluonCorr} still have a gauge link dependence. In the article dealing with the universality of the quark correlators~\cite{Buffing:2012sz} a similar situation occurred. In that paper, TMDs of a definite rank were introduced for quarks, which allowed the definition of universal quark TMDs, multiplied with gauge link dependent factors. In the same way as done for quarks, we will first show the effects of transverse weightings for the correlators at the operator level and in the next step use this to identify TMDs for the gluons.

\subsection{Operator structure of transverse moments\label{s:ftm}}
For gluons the single weighting $(m=1)$ results were already given in Ref.~\cite{Bomhof:2007xt} and for quarks the procedure was described in Ref.~\cite{Buffing:2012sz}, where also the decomposition of weighted matrix elements was given. As will be explained, the procedure for gluons is slightly more involved due to the larger number of possible color structures. 

\begin{figure}[t]
	\epsfig{file=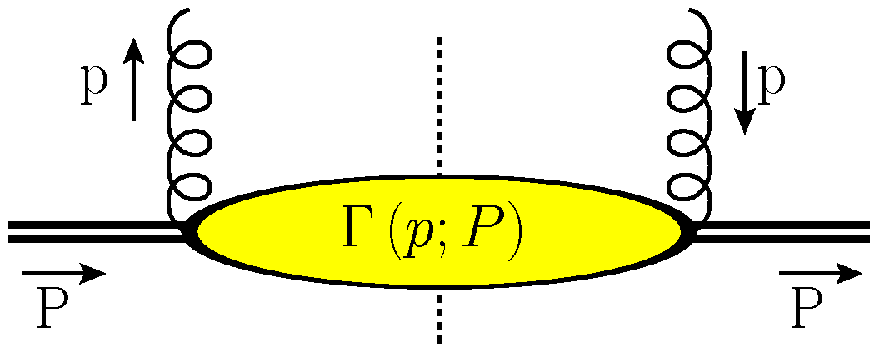,width=0.25\textwidth}
	\hspace{1.5cm}
	\epsfig{file=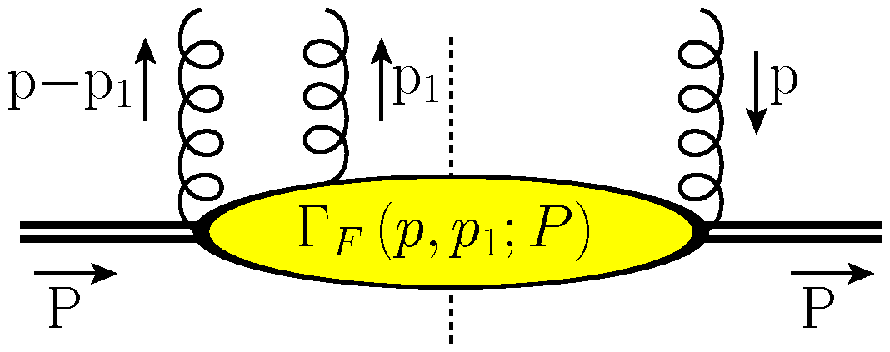,width=0.253388\textwidth}
	\\[0.2cm]
	(a)\hspace{58mm} (b)
\caption{\label{f:Phi_Acor}
The graphical representation of correlators 
(a) $\Gamma(p)$ and (b) $\Gamma_{F}(p,p-p_{1})$.} 
\end{figure}

Transverse weighting including transverse momenta is achieved in the transverse moments,
\bea
\Gamma_{\partial\ldots\partial}^{\alpha_1\ldots\alpha_m [U]}(x)
\equiv \int d^2p_{\st} 
\ p_\st^{\alpha_1}\ldots p_\st^{\alpha_m}\,\Gamma^{[U]}(x,p_\st) . 
\eea
In principle, a lot of different gauge link structures are possible, so it is important to understand the action of $p_{\st}^{\alpha}$ on the correlator $\Gamma^{[U]}(x,p_\st)$. In coordinate space the momentum $p_{\st}^{\alpha}$ becomes a partial derivative, which acts on the gauge links and yields~\cite{Buffing:2011mj,Buffing:2012sz}
\begin{subequations}
\begin{align}
&i\partial_\st^\alpha(\xi) U^{[\pm]}_{[0,\xi]}
= U^{[n]}_{[0,\pm\infty]}\,i\partial_\st^\alpha(\xi) U^\st_{[0_\st,\xi_\st]}
\,U^{[n]}_{[\pm\infty,\xi]}
= U^{[n]}_{[0,\pm\infty]}\,U^\st_{[0_\st,\xi_\st]}
\,iD_\st^\alpha(\pm\infty)\,U^{[n]}_{[\pm\infty,\xi]},
\\
&iD_{\st}^{\alpha}(\pm\infty)\,U_{[\pm\infty,\xi]}^{[n]}\ldots 
=U_{[\pm\infty,\xi]}^{[n]}\bigg(iD_{\st}^{\alpha}(\xi)-A_{\st}^{\alpha}(\xi)
\pm G_\st^{\alpha}(\xi)\bigg)\ldots \, , 
\\
&\ldots U_{[\xi,\pm\infty]}^{[n]}\,iD_{\st}^{\alpha}(\pm\infty)
= \ldots \bigg(iD_{\st}^{\alpha}(\xi)-A_{\st}^{\alpha}(\xi)\pm 
G_\st^{\alpha}(\xi)\bigg) U_{[\xi,\pm\infty]}^{[n]},
\label{e:TM2}
\end{align}
\end{subequations}
in which $A_{\st}^{\alpha}(\xi)$ and $G_\st^{\alpha}(\xi)$ are defined as
\bea
&&A_{\st}^{\alpha}(\xi)=\frac{1}{2}\int_{-\infty}^{\infty}
d\eta{\cdot}P\ \epsilon(\xi{\cdot}P-\eta{\cdot}P)
\,U_{[\xi,\eta]}^{[n]} F^{n\alpha}(\eta)U_{[\eta,\xi]}^{[n]}, \label{e:defA} \\
&&G_\st^{\alpha}(\xi)=\frac{1}{2}\int_{-\infty}^{\infty}
d\eta{\cdot}P\ U_{[\xi,\eta]}^{[n]}F^{n\alpha}(\eta)
U_{[\eta,\xi]}^{[n]}
\label{e:defG}
\eea
and where we have absorbed factors of {\it g} in the definition of the gluon fields. The latter operator combination only depends on $\xi_\st$. These specific field combinations have definite time-reversal properties, $A_\st(\xi)$ being T-even and $G_\st(\xi)$ being T-odd. Taking transverse moments thus requires the consideration of multi-parton matrix elements. The first transverse moment of a gluon correlator with a single color trace (type 1) involves after $p_\st$-integration the collinear correlators~\cite{Bomhof:2007xt}
\begin{subequations}
\begin{align}
&\Gamma_{D}^{\mu\nu,\alpha[U]}(x,x-x_1)=\int \frac{d\,\xi{\cdot}P}{2\pi}\frac{d\,\eta{\cdot}P}{2\pi}\ e^{ix_1 (\eta\cdot P)}e^{i (x-x_1)(\xi \cdot P)} \nonumber \\
&\mspace{260mu} \times\tr\langle P,S|F^{n\mu}(0)\big[U_{[0,\eta]}^{[n]}
iD_\st^{\alpha}(\eta)U_{[\eta,0]}^{[n]},U_{[0,\xi]}^{[n]}F^{n\nu}(\xi)U_{[\xi,0]}^{[n]}\big]|P,S\rangle\Big|_{\text{LC}}, 
\label{e:GammaD} \\
&\Gamma_{F,1}^{\mu\nu,\alpha[U]}(x,x-x_1)=\int \frac{d\,\xi{\cdot}P}{2\pi}\frac{d\,\eta{\cdot}P}{2\pi}\ e^{ix_1 (\eta\cdot P)}e^{i (x-x_1)(\xi \cdot P)} \nonumber \\
&\mspace{260mu} \times\tr\langle P,S|F^{n\mu}(0)\big[U_{[0,\eta]}^{[n]}
F^{n\alpha}(\eta)U_{[\eta,0]}^{[n]},U_{[0,\xi]}^{[n]}F^{n\nu}(\xi)U_{[\xi,0]}^{[n]}\big]|P,S\rangle\Big|_{\text{LC}}, 
\label{e:GammaF1} \\
&\Gamma_{F,2}^{\mu\nu,\alpha[U]}(x,x-x_1)=\int \frac{d\,\xi{\cdot}P}{2\pi}\frac{d\,\eta{\cdot}P}{2\pi}\ e^{ix_1 (\eta\cdot P)}e^{i (x-x_1)(\xi \cdot P)} \nonumber \\
&\mspace{260mu} \times\tr\langle P,S|F^{n\mu}(0)\big\{U_{[0,\eta]}^{[n]}
F^{n \alpha}(\eta)U_{[\eta,0]}^{[n]},U_{[0,\xi]}^{[n]}F^{n\nu}(\xi)U_{[\xi,0]}^{[n]}\big\}|P,S\rangle\Big|_{\text{LC}}, 
\label{e:GammaF2}
\end{align}
\end{subequations}
see Fig.~\ref{f:Phi_Acor} for the diagrammatic representation of $\Gamma_F(p,p-p_1)$.
For $\Gamma_D$ the color tracing involves the commutator. Note that in the operators we always use $i\partial$ and $iD$, whereas in indices of correlators we use $\partial$ and $D$. For $\Gamma_{F,c}$ one can have both the commutator and anticommutator, color structures that are distinguished using the indices $c=1$ and $c=2$, respectively. These collinear correlators have a unique gauge link structure. The actual matrix elements appearing in the transverse moments are bilocal, namely 
\begin{subequations}
\begin{align}
\Gamma_{D}^{\mu\nu,\alpha}(x)&=
\int dx_1\ \Gamma_{D}^{\mu\nu,\alpha}(x,x-x_1) \nonumber \\
&=\int \frac{d\,\xi{\cdot}P}{2\pi}\ e^{i x(\xi \cdot P)} \tr\langle P,S|F^{n\mu}
(0)U_{[0,\xi]}^{[n]}\big[iD_\st^{\alpha}(\xi),F^{n\nu}(\xi)\big] U_{[\xi,0]}^{[n]}
|P,S\rangle\Big|_{\text{LC}},
\label{e:PGammaD1} \\
\Gamma_A^{\mu\nu,\alpha}(x)&=\int dx_1\ \text{PV}\frac{i}{x_1}\Gamma_{F,1}^{\mu\nu,\alpha}(x,x-x_1) \nonumber\\
&=\int \frac{d\,\xi{\cdot}P}{2\pi}\ e^{i x(\xi \cdot P)} \tr\langle P,S|F^{n\mu}
(0)U_{[0,\xi]}^{[n]}\big[A_\st^{\alpha}(\xi),F^{n\nu}(\xi)\big] U_{[\xi,0]}^{[n]}
|P,S\rangle\Big|_{\text{LC}},
\label{e:PGammaA1} \\
\widetilde\Gamma_\partial^{\mu\nu,\alpha}(x) &\equiv
\Gamma_{D}^{\mu\nu,\alpha}(x)-\Gamma_{A}^{\mu\nu,\alpha}(x) \nonumber \\
&=\int \frac{d\,\xi{\cdot}P}{2\pi}\ e^{i x(\xi \cdot P)} \tr\langle P,S|F^{n\mu}
(0)U_{[0,\xi]}^{[n]}\big[i\partial_\st^{\alpha}(\xi),F^{n\nu}(\xi)\big] U_{[\xi,0]}^{[n]}
|P,S\rangle\Big|_{\text{LC}},
\label{e:PGammad1} \\
\Gamma_{G,1}^{\mu\nu,\alpha}(x)&=\Gamma_{F,1}^{\mu\nu,\alpha}(x,x)\nonumber\\
&=\int \frac{d\,\xi{\cdot}P}{2\pi}\ e^{i x(\xi \cdot P)} \tr\langle P,S|F^{n\mu}
(0)U_{[0,\xi]}^{[n]}\big[G_\st^{\alpha}(\xi),F^{n\nu}(\xi)\big] U_{[\xi,0]}^{[n]}
|P,S\rangle\Big|_{\text{LC}},
\label{e:PGammaG1} \\
\Gamma_{G,2}^{\mu\nu,\alpha}(x)&=\Gamma_{F,2}^{\mu\nu,\alpha}(x,x) \nonumber\\
&=\int \frac{d\,\xi{\cdot}P}{2\pi}\ e^{i x(\xi \cdot P)} \tr\langle P,S|F^{n\mu}
(0)U_{[0,\xi]}^{[n]}\bigl\{G_\st^{\alpha}(\xi),F^{n\nu}(\xi)\bigr\} U_{[\xi,0]}^{[n]}
|P,S\rangle\Big|_{\text{LC}}.
\label{e:PGammaG2}
\end{align}
\end{subequations}
The matrix elements needed in the first moment thus involve operator structures $F^{n\mu}(0)\,U_{[0,\xi]}^{[n]}O(\xi)\,F^{n\nu}(\xi)U_{[\xi,0]}^{[n]}$, with $O$ being $iD_\st^\alpha$, $A_\st^\alpha$, $i\partial_\st^\alpha = iD_\st^\alpha - A_\st^\alpha$ or $G_\st^{\alpha}$ and involving for $\Gamma_G$ multiple color tracing possibilities. For $\Gamma_D (x)$, $\Gamma_A (x)$ and their difference $\widetilde\Gamma_\partial(x)$ there is only one color structure. For higher moments we also need bilocal correlators $\widetilde\Gamma_{\partial\ldots\partial}^{\mu\nu,\alpha_1\ldots\alpha_m}$, $\Gamma_{G\ldots G,c}^{\mu\nu,\alpha_1\ldots\alpha_m}$ and symmetrized combinations $\widetilde\Gamma_{\{\partial\ldots\partial G\ldots G\},c}^{\mu\nu,\alpha_1\ldots\alpha_m}$, again with index $c$ to distinguish multiple color configurations. These operator structures will be discussed in the next section. We refer to the number of gluonic pole contributions in these bilocal correlators as the gluonic pole rank and to the sum of gluonic poles and partial derivative terms as the rank of the correlator, which equals the number of transverse indices.

In order to obtain the color structure for higher moments, one needs to carefully look at the results for the derivative before $p_\st$-integration. The basic results that will be needed are
\begin{subequations}
\begin{align}
&\left[i\partial_\st^\alpha(\xi), U_{[0,\xi]}^{[\pm]}\right] =
\pm U_{[0,\xi]}^{[\pm]}\,G_\st^{\alpha}(\xi)
\quad\mbox{and}\quad
\left[i\partial_\st^\alpha(\xi), U_{[\xi,0]}^{[\mp]}\right] =
\pm G_\st^{\alpha}(\xi)\,U_{[\xi,0]}^{[\mp]}, \,
\label{e:derivativeslink1a} \\
&\left[i\partial_\st^\alpha(\xi), U^{[\Box]}\right] =
2\,U_{[0,\xi]}^{[+]}\,G_\st^{\alpha}(\xi)U_{[\xi,0]}^{[-]}
\quad\mbox{and}\quad
\left[i\partial_\st^\alpha(\xi), U^{[\Box]\dagger}\right] =
-2\,U_{[0,\xi]}^{[-]}\,G_\st^{\alpha}(\xi)\,U_{[\xi,0]}^{[+]}.
\label{e:derivativeslink1b}
\end{align}
\label{e:derivativeslink1}
\end{subequations}
Since we can also write $U^{[\Box]} = U^{[-]}_{[\xi,0]}U^{[+]}_{[0,\xi]}$, one also has
\begin{equation}
\left[i\partial_\st^\alpha(\xi), U^{[\Box]}\right] =
\left\{G_\st^{\alpha}(\xi),U^{[\Box]}\right\}
\quad\mbox{and}\quad 
\left[i\partial_\st^\alpha(\xi), U^{[\Box]\dagger}\right] =
-\left\{G_\st^{\alpha}(\xi),U^{[\Box]\dagger}\right\} .
\label{e:derivativeslink2}
\end{equation}
To get the results for expressions including field tensors or get to higher derivatives, we can use recursive relations,
\begin{subequations}
\begin{align}
&\left[i\partial_\st^{\alpha_1}(\xi), U_{[0,\xi]}^{[\pm]}\,O(\xi)\right] =
U_{[0,\xi]}^{[\pm]}\,\left[i\partial_\st^{\alpha_1}(\xi), O(\xi)\right]
\pm U_{[0,\xi]}^{[\pm]}\,G_\st^{\alpha_1}(\xi)\,O(\xi),
\label{e:derivativeslink3a} 
\\&
\left[i\partial_\st^{\alpha_1}(\xi), 
U_{[0,\xi]}^{[\pm]}\,O(\xi)\,U_{[\xi,0]}^{[\pm]}\right] =
U_{[0,\xi]}^{[\pm]}\big[i\partial_\st^{\alpha_1}(\xi),O(\xi)\big]
U_{[\xi,0]}^{[\pm]} \pm
U_{[0,\xi]}^{[\pm]}\big[G_\st^{\alpha_1}(\xi),O(\xi)\big]
U_{[\xi,0]}^{[\pm]} ,
\label{e:derivativeslink3b}
\\&
\left[i\partial_\st^{\alpha_1}(\xi), 
U_{[0,\xi]}^{[\pm]}\,O(\xi)\,U_{[\xi,0]}^{[\mp]}\right] =
U_{[0,\xi]}^{[\pm]}\big[i\partial_\st^{\alpha_1}(\xi),O(\xi)\big]
U_{[\xi,0]}^{[\mp]} \pm
U_{[0,\xi]}^{[\pm]}\big\{G_\st^{\alpha_1}(\xi),O(\xi)\big\}
U_{[\xi,0]}^{[\mp]} ,
\label{e:derivativeslink3c}
\end{align}
\label{e:derivativeslink3}
\end{subequations}
where $O(\xi)$ can be field strengths $F^{n\alpha_2}(\xi)$, but also expressions like $[i\partial_\st^{\alpha_2},F^{n\alpha_3}(\xi)]$, $G_\st^{\alpha_2}(\xi)$ or $U^{[\Box]}$ as well as commutators or anticommutators of these such as $[G_\st^{\alpha_2}(\xi),F^{n\alpha_3}(\xi)]$, $\{G_\st^{\alpha_2}(\xi),U^{[\Box]}\}$, etc. In all (unweighted) bilocal matrix elements including gauge link structures the $\xi$-dependent part within a particular trace can be written in the form $U^{[\pm]}O\left(\xi\right)U^{[\pm]}$ and $U^{[\pm]}O\left(\xi\right)U^{[\mp]}$. In this, the operator $O\left(\xi\right)$ can be the unit operator, a gluon field $F(\xi)$, a gluon field with an additional gauge link attached, e.g. $U^{[\square]}F(\xi)$, etc. Taking a transverse moment implies taking a transverse derivative on the above combinations of gauge links and fields. For a single transverse weighting this results in an (anti)commutator of a gluonic pole term with the operator structure $O$ and the transverse derivative of the operator structure $O$ as outlined above. For a double transverse weighting one has to apply this to the result of the single weighted result, etc. The results of this are tabulated in Table~\ref{t:operators} and will be used to calculate higher moments.
\begin{table}[!tb]
\centering
\begin{tabular}{p{12mm}|p{67mm}|p{67mm}}
Rank
&$U^{[\pm]}_{[0,\xi]}O\left(\xi\right)U_{[\xi,0]}^{[\pm]}$
&$U_{[0,\xi]}^{[\pm]}O\left(\xi\right)U_{[\xi,0]}^{[\mp]}$
\\[2pt]
\hline
\hline
$m=0$
&$O$
&$O$
\\[5pt]
$m=1$
&$O^{\prime}\pm\big[G,O\big]$
&$O^{\prime}\pm\big\{G,O\big\}$
\\[5pt]
$m=2$
&$O^{\prime\prime}\pm\big[G^\prime,O\big]\pm 2\big[G,O^\prime\big]+\big[G,\big[G,O\big]\big]$
&$O^{\prime\prime}\pm\big\{G^\prime,O\big\}\pm 2\big\{G,O^\prime\big\}+\big\{G,\big\{G,O\big\}\big\}$
\\[5pt]
$m=3$
&$O^{\prime\prime\prime}\pm\big[G^{\prime\prime},O\big]\pm 3\big[G^{\prime},O^{\prime}\big]\pm 3\big[G,O^{\prime\prime}\big]$ \newline
$+\big[G^{\prime}\big[G,O\big]\big]+2\big[G,\big[G^{\prime},O\big]\big]+3\big[G\big[G,O^{\prime}\big]\big]$\newline
$\pm\big[G\big[G\big[G,O\big]\big]\big]$
&$O^{\prime\prime\prime}\pm\big\{G^{\prime\prime},O\big\}\pm 3\big\{G^{\prime},O^{\prime}\big\}\pm 3\big\{G,O^{\prime\prime}\big\}$ \newline
$+\big\{G^{\prime}\big\{G,O\big\}\big\}+2\big\{G,\big\{G^{\prime},O\big\}\big\}+3\big\{G\big\{G,O^{\prime}\big\}\big\}$ \newline
$\pm\big\{G\big\{G\big\{G,O\big\}\big\}\big\}$
\\[2pt]
\hline
\end{tabular}
\parbox{0.85\textwidth}{
\caption{\label{t:operators}
Transverse derivatives of the two basic color structures for a given rank $m$ corresponding to taking $m$ transverse derivatives $i\partial_\st^{\alpha_1}\ldots i\partial_\st^{\alpha_m}$. The primes denote transverse partial derivatives of the operators $O$ or $G$, $[i\partial_\st^\alpha,O] \rightarrow O^\prime$ and $[i\partial_\st^\alpha,G_\st^{\beta}] \rightarrow G^\prime$. Since taking the derivatives is symmetric, one needs to make sure that when the transverse indices are made explicit, they are symmetrized and averaged. Furthermore the gauge links $U^{[\pm]}\ldots U^{[\pm]}$ or $U^{[\pm]}\ldots U^{[\mp]}$ need to be added. Eqs.~\ref{e:derivativeslink3b} and \ref{e:derivativeslink3c} constitute the $m = 1$ results including indices.}}
\end{table}

To calculate transverse moments one needs to perform the $p_\st$-integrations. In those results the gauge links $U^{[\pm]}$ will reduce to straight-line gauge links $U^{[n]}$ along the light-like direction $n$. A Wilson loop will reduce to the identity in color space after such an integration, implying also
\begin{equation} 
\int d^2 p_\st \ldots\left\{G_\st^{\alpha}(\xi),U^{[\Box]}\right\}\ldots\rightarrow \ldots 2\,G_\st^{\alpha}(\xi)\ldots
\quad\mbox{and}\quad
\int d^2 p_\st \ldots\left\{G_\st^{\alpha}(\xi),U^{[\Box]\dagger}\right\}\ldots\rightarrow \ldots 2\,G_\st^{\alpha}(\xi)\ldots
\label{e:derivativeslinkmain}
\end{equation}
in the collinear (hence integrated) situation. As an explicit example, consider $F(0)U^{[+\Box]}F(\xi)U^{[+\Box]\dagger}$. In that case one applies the transverse derivative to $O(\xi) = U^{[\Box]}F(\xi)U^{[\Box]\dagger}$ sandwiched between $U^{[+]}\ldots U^{[+]\dagger}$, yielding according to Eq.~\ref{e:derivativeslink3b} the combination $[i\partial_\st + G_\st,U^{[\Box]}FU^{[\Box]\dagger}]$. This can be evaluated using the basic results for gauge links, specifically Eq.~\ref{e:derivativeslink2}, resulting after $p_\st$-integration into the operator $[i\partial_\st,F] + 3\,[G_\st,F]$ sandwiched between $U^{[n]}$ links. This factor 3 multiplying $\Gamma_G$ in the transverse moment is an explicit example of what in general is referred to as gluonic pole factors (see next subsection). Note that for higher moments it sometimes will be convenient to include some factors of two in the correlator definitions (outlined in section~\ref{s:colorstructures}) rather than using anticommutators.

\subsection{Example: single transverse moments}
Using the previously given relations for taking derivatives of Wilson lines, one can calculate the transverse moments, i.e.\ the matrix elements including transverse momentum weightings and integrated over $p_\st$. As an example, we will give the explicit results for the single weighted case of a gluon correlator with a single color trace (type 1). It is found that
\begin{eqnarray}
\Gamma_{\partial}^{\alpha [U]}(x)&\equiv& \int d^2p_{\st}
\ p_{\st}^{\alpha}\,\Gamma^{[U]}(x,p_\st) \nonumber \\
&=&\Big(\Gamma_{D}^{\alpha}(x)-\Gamma_A^{\alpha}(x)\Big)
+C_{G,1}^{[U]}\,\Gamma_{G,1}^{\alpha}(x)
+C_{G,2}^{[U]}\,\Gamma_{G,2}^{\alpha}(x)
\nonumber \\
&=&\widetilde\Gamma_{\partial}^{\alpha}(x)
+C_{G,1}^{[U]}\,\Gamma_{G,1}^{\alpha}(x)
+C_{G,2}^{[U]}\,\Gamma_{G,2}^{\alpha}(x).
\label{e:Phiw1-1}
\end{eqnarray}
In the above procedure, one finds universal matrix elements and all process dependence is isolated in gluonic pole factors $C_{G,c}^{[U]}$. The numerical values of these coefficients for all different gauge links can be found in the Tables~\ref{t:gpfactors1}, \ref{t:gpfactors2} and \ref{t:gpfactors3}. For example, as explicitly outlined in the previous subsection, one finds the factor $C_{G,1}^{[+\Box ,+^\dagger\Box^\dagger}] = +3$. For the single weighted case they are related to the color factors given in Ref.~\cite{Qiu:1998ia}. In Eq.~\ref{e:Phiw1-1}, our $C_{G,1}^{[U]}\,\Gamma_{G,1}^{\alpha[U]}(x)$ and $C_{G,2}^{[U]}\,\Gamma_{G,2}^{\alpha[U]}(x)$ are the same as the $\pi C_{G_f}^{[U]}\,\Gamma_{G_f}^{\alpha[U]}(x,x)$ and $\pi C_{G_d}^{[U]}\,\Gamma_{G_d}^{\alpha[U]}(x,x)$ in Ref.~\cite{Bomhof:2007xt}.

\section{Operator structures of all transverse moments\label{s:colorstructures}}
In this section we give the results including the transverse moments of rank higher than one, which was done in detail in the previous section. The necessary ingredients are finding the operators containing derivatives and gluonic poles and the labeling of the color structures for the operators that contain gluonic poles. Having done that, one can find the factors in the expansion of the transverse moments. Following up on the result for the first transverse moment in Eq.~\ref{e:Phiw1-1} we have
\begin{eqnarray}
\Gamma_{\partial}^{\alpha_1\,[U]}(x)
&\equiv& \int d^2p_{\st} \,p_{\st}^{\alpha_1}
\,\Gamma^{[U]}(x,p_\st) \nonumber \\
&=&\widetilde\Gamma_{\partial}^{\alpha_1}(x)
+ \sum_c C_{G,c}^{[U]}\,\Gamma_{G,c}^{\alpha_1}(x),
\label{e:Phiw1} \\
\Gamma_{\partial\partial}^{\alpha_1\alpha_2\,[U]}(x)
&\equiv& \int d^2p_{\st} \,p_{\st}^{\alpha_1}p_{\st}^{\alpha_2}
\,\Gamma^{[U]}(x,p_\st) \nonumber \\
&=&\widetilde\Gamma_{\partial\partial}^{\alpha_1\alpha_2}(x)
+ \sum_c C_{G,c}^{[U]}
\,\widetilde\Gamma_{\{\partial G\},c}^{\alpha_1\alpha_2}(x)
+ \sum_c C_{GG,c}^{[U]}\,\Gamma_{GG,c}^{\alpha_1\alpha_2}(x),
\label{e:Phiw2} \\
\Gamma_{\partial\partial\partial}^{\alpha_1\alpha_2\alpha_3\,[U]}(x)
&\equiv& \int d^2p_\st\ p_{\st}^{\alpha_1}p_\st^{\alpha_2}p_\st^{\alpha_3}
\,\Gamma^{[U]}(x,p_\st) \nonumber \\
&=&\widetilde\Gamma_{\partial\partial\partial}^{\alpha_1\alpha_2\alpha_3}(x)
+ \sum_c C_{G,c}^{[U]}
\,\widetilde\Gamma_{\{\partial\partial G\},c}^{\alpha_1\alpha_2\alpha_3}(x)
+ \sum_c C_{GG,c}^{[U]}
\,\widetilde\Gamma_{\{\partial GG\},c}^{\alpha_1\alpha_2\alpha_3}(x)
+ \sum_c C_{GGG,c}^{[U]}\,\Gamma_{GGG,c}^{\alpha_1\alpha_2\alpha_3}(x),
\label{e:Phiw3}
\end{eqnarray}
where $[U]$ can be any of the gauge links discussed in section~\ref{s:gaugelinks}. The collinear correlators on the rhs are independent of the gauge link. The gauge link dependence is in the gluonic pole factors multiplying these correlators. It should be noted that the coefficients $C_{G,c}^{[U]}$ for a given gauge link are the same in the expressions for single, double and triple transverse moments, as are also the coefficients $C_{GG,c}^{[U]}$ in double and triple transverse moments. They do not depend on the number of partial derivatives involved. The results have been tabulated for type 1 correlators in Table~\ref{t:gpfactors1}, for type 2 correlators in Table~\ref{t:gpfactors2} and for type 3 correlators in Table~\ref{t:gpfactors3}. Note that not all the gauge links in the Tables~\ref{t:gpfactors1} and \ref{t:gpfactors2} will occur for correlators in actual $2\rightarrow 2$ diagrams, but they are needed for the description in Appendix~\ref{A:operator}. The color structures for these types are explicitly given in the next sections. Also the gluonic pole coefficients for the single weighted case for $2\rightarrow 2$ processes are tabulated in Ref.~\cite{Bomhof:2006ra}.

\begin{table}[!tb]
\begin{tabular}{|l||>{\centering}m{11mm} >{\centering}m{11mm} | >{\centering}m{11mm} >{\centering}m{11mm} | c c|}
\hline
$\rule{0pt}{4mm}\Gamma^{[U]}$
&$C_{G,1}^{[U]}$
&$C_{G,2}^{[U]}$
&$C_{GG,1}^{[U]}$
&$C_{GG,2}^{[U]}$
&$C_{GGG,1}^{[U]}$
&$C_{GGG,2}^{[U]}$
\tabularnewline[2pt]
\hline
\hline
$\Gamma^{[+,+^\dagger]}$&$1$&$0$&$1$&$0$&$1$&$0$
\tabularnewline[1pt]
$\Gamma^{[-,-^\dagger]}$&$-1$&$0$&$1$&$0$&$-1$&$0$
\tabularnewline[1pt]
\hline
$\Gamma^{[+,-^\dagger]}$&$0$&$1$&$0$&$1$&$0$&$1$
\tabularnewline[1pt]
$\Gamma^{[-,+^\dagger]}$&$0$&$-1$&$0$&$1$&$0$&$-1$
\tabularnewline[1pt]
\hline
$\Gamma^{[+\square,+^\dagger\square^\dagger]}$&$3$&$0$&$9$&$0$&$27$&$0$
\tabularnewline[1pt]
$\Gamma^{[-\square^\dagger,-^\dagger\square]}$&$-3$&$0$&$9$&$0$&$-27$&$0$
\tabularnewline[1pt]
\hline
$\Gamma^{[+\square,-^\dagger\square]}$&$0$&$3$&$0$&$9$&$0$&$27$
\tabularnewline[1pt]
$\Gamma^{[-\square^\dagger,+^\dagger\square^\dagger]}$&$0$&$-3$&$0$&$9$&$0$&$-27$
\tabularnewline[1pt]\hline
\end{tabular}
\parbox{0.85\textwidth}{
\caption{
The values of the gluonic pole factors for the type 1 gauge links needed in the $p_{\st}$-weighted cases. Note that the value of $C_{G,1}^{[U]}$, $C_{G,2}^{[U]}$, $C_{GG,1}^{[U]}$ and $C_{GG,2}^{[U]}$ are the same for single, double and triple transverse weighting. The coefficients that are not present in the above table are zero for these gauge links.
\label{t:gpfactors1}}}\\[4mm]
\begin{tabular}{|l||>{\centering}m{11mm} >{\centering}m{11mm} | >{\centering}m{11mm} >{\centering}m{11mm} >{\centering}m{11mm} | c c c c c|}
\hline
$\rule{0pt}{4mm}\Gamma^{[U]}$
&$C_{G,1}^{[U]}$
&$C_{G,2}^{[U]}$
&$C_{GG,1}^{[U]}$
&$C_{GG,2}^{[U]}$
&$C_{GG,3}^{[U]}$
&$C_{GGG,1}^{[U]}$
&$C_{GGG,2}^{[U]}$
&$C_{GGG,3}^{[U]}$
&\centering $C_{GGG,4}^{[U]}$
&$C_{GGG,5}^{[U]}$
\tabularnewline[2pt]
\hline
\hline
$\Gamma^{[+,+^{\dagger}(\square)]}$&
$1$&$0$&$1$&$0$&$1$&$1$&$0$&$3$&$0$&$1$
\tabularnewline[1pt]
$\Gamma^{[-,-^{\dagger}(\square^\dagger)]}$&
$-1$&$0$&$1$&$0$&$1$&$-1$&$0$&$-3$&$0$&$-1$
\tabularnewline[1pt]
\hline
$\Gamma^{[+,+^{\dagger}(\square^{\dagger})]}$&
$1$&$0$&$1$&$0$&$1$&$1$&$0$&$3$&$0$&$-1$
\tabularnewline[1pt]
$\Gamma^{[-,-^{\dagger}(\square)]}$&
$-1$&$0$&$1$&$0$&$1$&$-1$&$0$&$-3$&$0$&$1$
\tabularnewline[1pt]
\hline
$\Gamma^{[+,-^{\dagger}(\square^{\dagger})]}$&
$0$&$1$&$0$&$1$&$1$&$0$&$1$&$0$&$3$&$-1$
\tabularnewline[1pt]
$\Gamma^{[-,+^{\dagger}(\square)]}$&
$0$&$-1$&$0$&$1$&$1$&$0$&$-1$&$0$&$-3$&$1$
\tabularnewline[1pt]
\hline
$\Gamma^{[+,+^{\dagger}(\square)(\square^{\dagger})]}$&
$1$&$0$&$1$&$0$&$2$&$1$&$0$&$6$&$0$&$0$
\tabularnewline[1pt]
$\Gamma^{[-,-^{\dagger}(\square)(\square^{\dagger})]}$&
$-1$&$0$&$1$&$0$&$2$&$-1$&$0$&$-6$&$0$&$0$
\tabularnewline[1pt]
\hline
$\Gamma^{[+,-^{\dagger}(\square)(\square^{\dagger})]}$&
$0$&$1$&$0$&$1$&$2$&$0$&$1$&$0$&$6$&$0$
\tabularnewline[1pt]
$\Gamma^{[-,+^{\dagger}(\square)(\square^{\dagger})]}$&
$0$&$-1$&$0$&$1$&$2$&$0$&$-1$&$0$&$-6$&$0$
\tabularnewline[1pt]
\hline
\end{tabular}
\parbox{0.85\textwidth}{
\caption{
The values of the gluonic pole factors for gauge links needed in the $p_{\st}$-weighted cases containing (traced) Wilson loops. The coefficients that are not present in the above table are zero for these gauge links.
\label{t:gpfactors2}}}\\[4mm]
\begin{tabular}{|l||>{\centering}m{11mm} | c c|}
\hline
$\rule{0pt}{4mm}\Gamma^{[U]}$
&$C_{GG,4}^{[U]}$
&$C_{GGG,6}^{[U]}$
&$C_{GGG,7}^{[U]}$
\tabularnewline[2pt]
\hline
\hline
$\Gamma^{[(F(\xi)\square),(F(0)\square^{\dagger})]}$&
$-2$&$3$&$-3$
\tabularnewline[1pt]
$\Gamma^{[(F(\xi)\square^{\dagger}),(F(0)\square)]}$&
$-2$&$-3$&$3$
\tabularnewline[1pt]
\hline
\end{tabular}
\parbox{0.85\textwidth}{
\caption{
The values of the gluonic pole factors for gauge links needed in the $p_{\st}$-weighted cases containing two traced gluon fields combined with Wilson loops. The coefficients that are not present in the above table are zero for these gauge links.
\label{t:gpfactors3}}}
\end{table}

\subsection{Weighting for type 1 correlators}
Unweighted correlators of type 1 have the field theoretical structure given in Eq.~\ref{e:type1}. Applying transverse weightings on this type of matrix elements implies that all gluonic pole matrix elements that appear are located in the same color trace as the gluon fields $F(0)$ and $F(\xi)$. If one considers matrix elements containing gluonic poles only and no partial derivative terms, one gets for single, double and triple transverse weighting the matrix elements
\begin{subequations}
\begin{align}
\Gamma_{G,1}^{\alpha_1}&\rightarrow \tr_c\Big\lgroup F(0) \left[G_\st^{\alpha_1}(\xi),F(\xi)\right] \Big\rgroup,
\label{e:colorG1}\\
\Gamma_{G,2}^{\alpha_1}&\rightarrow \tr_c\Big\lgroup F(0) \left\{G_\st^{\alpha_1}(\xi),F(\xi)\right\} \Big\rgroup,
\label{e:colorG2}\\
\Gamma_{GG,1}^{\alpha_1\alpha_2}&\rightarrow \tr_c\Big\lgroup F(0) \left[G_\st^{\alpha_1}(\xi),\left[G_\st^{\alpha_2}(\xi),F(\xi)\right]\right]\Big\rgroup,
\label{e:colorGG1}\\
\Gamma_{GG,2}^{\alpha_1\alpha_2}&\rightarrow \tr_c\Big\lgroup F(0) \left\{G_\st^{\alpha_1}(\xi),\left\{G_\st^{\alpha_2}(\xi),F(\xi)\right\}\right\} \Big\rgroup,
\label{e:colorGG2}\\
\Gamma_{GGG,1}^{\alpha_1\alpha_2\alpha_3}&\rightarrow \tr_c\Big\lgroup F(0) \left[G_\st^{\alpha_1}(\xi),\left[G_\st^{\alpha_2}(\xi),\left[G_\st^{\alpha_3}(\xi),F(\xi)\right]\right]\right] \Big\rgroup,
\label{e:colorGGG1}\\
\Gamma_{GGG,2}^{\alpha_1\alpha_2\alpha_3}&\rightarrow \tr_c\Big\lgroup F(0) \left\{G_\st^{\alpha_1}(\xi),\left\{G_\st^{\alpha_2}(\xi),\left\{G_\st^{\alpha_3}(\xi),F(\xi)\right\}\right\}\right\} \Big\rgroup,
\label{e:colorGGG2}
\end{align}
\end{subequations}
where we omitted the (collinear) gauge links for readability. As was indicated for the single weighted case already in the Refs.~\cite{Bomhof:2007xt,Dominguez:2010xd}, the gluonic pole terms come either in the form of a commutator or anticommutator with the field $F(\xi)$, depending on the gauge link structure. Since gluonic pole matrix elements are obtained in constructing transverse moments of a particular $O(\xi)$ operator combination sandwiched between two basic links, one finds that the matrix elements containing just multiple gluonic poles have only commutators or only anticommutators of the gluonic pole terms.

As pointed out before in the description of the formalism, also matrix elements containing partial derivative contributions appear. For the single weighted case this has been illustrated already. For double and triple weightings the situation becomes more involved, since terms containing both gluonic poles and partial derivative contributions will appear. The partial derivative term will always come as a commutator with the field $F(\xi)$. Again omitting the (collinear) gauge links, we get matrix elements like
\begin{subequations}
\begin{align}
\widetilde\Gamma_{\partial}^{\alpha_1}&\rightarrow \tr_c\Big\lgroup F(0) \left[i\partial_{\st}^{\alpha_1},F(\xi)\big]\right] \Big\rgroup, \\
\widetilde\Gamma_{\partial\partial}^{\alpha_1\alpha_2}&\rightarrow \tr_c\Big\lgroup F(0) \big[i\partial_{\st}^{\alpha_1}(\xi),\big[i\partial_{\st}^{\alpha_2}(\xi),F(\xi)\big]\big] \Big\rgroup, \\
\widetilde\Gamma_{\partial\partial\partial}^{\alpha_1\alpha_2\alpha_3}&\rightarrow \tr_c\Big\lgroup F(0) \big[i\partial_{\st}^{\alpha_1}(\xi),\big[i\partial_{\st}^{\alpha_2}(\xi),\big[i\partial_{\st}^{\alpha_3}(\xi),F(\xi)\big]\big]\big] \Big\rgroup . 
\end{align}
\end{subequations}
The matrix elements with only gluonic poles or with only derivatives are evidently symmetric for both sides of the expression. For double transverse weighting, one gets in addition mixed matrix elements. Looking at the results in Table~\ref{t:operators} we note that
\begin{eqnarray*} 
[G^\prime,O] + 2[G,O^\prime] &=& [\partial,[G,O]]+[G,[\partial,O]], \\
\{G^\prime,O\} + 2\{G,O^\prime\} &=& [\partial,\{G,O\}]+\{G,[\partial,O]\},
\end{eqnarray*}
which implies two types of mixed rank 2 operator structures appearing in the transverse moments
\begin{subequations}
\begin{align}
\widetilde\Gamma_{\{\partial G\},1}^{\{\alpha_1\alpha_2\}}&\rightarrow \tr_c\Big\lgroup 
F(0) \big[i\partial_\st^{\{\alpha_1},\big[G_\st^{\alpha_2\}},F(\xi)\big]\big] + F(0) \big[G_\st^{\{\alpha_1},\big[i\partial_{\st}^{\alpha_2\}},F(\xi)\big]\big] \Big\rgroup, \\
\widetilde\Gamma_{\{\partial G\},2}^{\{\alpha_1\alpha_2\}}&\rightarrow \tr_c\Big\lgroup F(0) \big[i\partial_{\st}^{\{\alpha_1},\big\{G_\st^{\alpha_2\}},F(\xi)\big\}\big] + F(0) \big\{G_\st^{\{\alpha_1},\big[i\partial_{\st}^{\alpha_2\}},F(\xi)\big]\big\} \Big\rgroup,
\end{align}
\end{subequations}
where we have suppressed the arguments ($\xi$) of $\partial_\st$ and $G_\st$ and $\{\partial G\}$ and $\{\alpha_1 \ldots \alpha_2\}$ denote symmetrization. This makes for combinations involving commutators and anticommutators the rhs evidently symmetric. By symmetrizing also the lhs we avoid additional numerical factors. Note that for the rank 3 mixed operator combinations
\begin{eqnarray*}
[G^{\prime\prime},O] + 3[G^\prime,O^\prime]+ 3[G,O^{\prime\prime}] &=& 
[\partial,[\partial,[G,O]]]+[\partial,[G,[\partial,O]]] + [G,[\partial,[\partial, O]]], \\
\{G^{\prime\prime},O\} + 3\{G^\prime,O^\prime\}+ 3\{G,O^{\prime\prime}\} &=& 
[\partial,[\partial,\{G,O\}]]+[\partial,\{G,[\partial,O]\}] + \{G,[\partial,[\partial, O]]\}
\end{eqnarray*}
and 
\begin{eqnarray*}
[G^\prime,[G,O]] + 2[G,[G^\prime,O]]+ 3[G,[G,O^\prime]] &=& 
[\partial,[G,[G,O]]]+[G,[\partial,[G,O]]] + [G,[G,[\partial, O]]], \\
\{G^\prime,\{G,O\}\} + 2\{G,\{G^\prime,O\}\}+ 3\{G,\{G,O^\prime\}\} &=& 
[\partial,\{G,\{G,O\}\}]+\{G,[\partial,\{G,O\}]\} + \{G,\{G,[\partial, O]\}\},
\end{eqnarray*}
we see the natural appearance of correlators with a symmetric operator structure, $\Gamma_{\{\partial\partial G\},c}$ and $\Gamma_{\{\partial GG\},c}$, each of these in two color configurations that just depend on the number of gluonic poles in the correlator (compare Eqs.~\ref{e:colorG1}, \ref{e:colorG2} and Eqs.~\ref{e:colorGG1}, \ref{e:colorGG2}). It should be noted that the coefficient $C_{G,c}^{[U]}$ is the same for single and double transverse weighting, thus the matrix elements $\Gamma_{G,c}$ and $\Gamma_{\{\partial G\},c}$ are multiplied with the same gluonic pole factor. Similarly, for triple transverse weighting one gets the additional mixed matrix elements $\Gamma_{\{\partial\partial G\},c}$ and $\Gamma_{\{\partial GG\},c}$, whose gluonic pole factors are identical as the gluonic pole factors of $\Gamma_{G,c}$ and $\Gamma_{GG,c}$ respectively. Note that in expressions involving transverse weightings, there should be symmetrization over the indices $\alpha_i$ and these expressions should be traceless. This applies to all correlators involving double or higher transverse weighting.

\subsection{Weighting for type 0 correlators}
Before turning to the type 2 and type 3 gluon correlators, we will discuss the weighting for type 0 correlators given by Eq.~\ref{e:type0} first, using the formalism in section~\ref{s:ftm}. To make things specific, we consider a diffractive correlator of the form $\Gamma_0^{[(\Box)]}(p_\st;n)$ containing a matrix element of the operator $\tfrac{1}{N_c}\,\tr_c\left\lgroup U^{[\text{loop}]}(\xi)-1\right\rgroup$. Using the results in section~\ref{s:ftm} one finds that (subtracting the unit operator) the $p_\st$-integrated result and the first transverse moment are both zero. The first nonzero contributions come at rank 2 (see also Ref.~\cite{Dominguez:2011wm}), a result that can be obtained from Table~\ref{t:operators}, e.g.\ choosing $O(\xi) = 1$ for $U^{[\text{loop}]} = U^{[(\Box)]}$, 
\begin{equation}
\Gamma_{0\ GG}^{\alpha_1\alpha_2}\rightarrow \tfrac{1}{N_c}\tr_c\Big\lgroup 
\left\{G_\st^{\alpha_1},\left\{G_\st^{\alpha_2},1\right\}\right\} \Big\rgroup
=\tfrac{2}{N_c}\tr_c\Big\lgroup \left\{G_\st^{\alpha_1}, G_\st^{\alpha_2}\right\}\Big\rgroup 
=\tfrac{4}{N_c}\tr_c\Big\lgroup G_\st^{\alpha_1} G_\st^{\alpha_2} \Big\rgroup.
\label{e:colorGG0} 
\end{equation}
At rank 3 one encounters the following collinear operator structures,
\begin{subequations}
\begin{align}
\widetilde\Gamma_{0\ \{\partial GG\}}^{\{\alpha_1\alpha_2\alpha_3\}}&\rightarrow 
\tfrac{2}{N_c}\tr_c\Big\lgroup 
\big[i\partial_\st^{\{\alpha_1},\big\{G_\st^{\alpha_2},G_\st^{\alpha_3\}}\big\}\big]
+ \big\{ G_\st^{\{\alpha_1},\big[i\partial_\st^{\alpha_2},G_\st^{\alpha_3\}}\big]\big\}\Big\rgroup ,
\\
\Gamma_{0\ GGG}^{\alpha_1\alpha_2\alpha_3}&\rightarrow 
\tfrac{2}{N_c}\tr_c\Big\lgroup 
\big\{G_\st^{\alpha_1},\big\{G_\st^{\alpha_2},G_\st^{\alpha_3}\big\}\big\} \Big\rgroup .
\end{align}
\end{subequations}
Even in more complex situations the Wilson loop reduces to unity after $p_\st$-integration and the same collinear operator structures are obtained, multiplied with specific gluonic pole factors in the actual transverse moments.

\subsection{Weighting for type 2 correlators}
Unweighted correlators of type 2 have the field theoretical structure given in Eq.~\ref{e:type2}. The weighting with factors of $p_{\st}$ acts on gluon fields and gauge links that depend on the coordinate $\xi_{\st}$. The matrix elements of the type 2 correlator contain one or more additional color traced Wilson loops, as a result of which the gluonic pole terms could appear in the color trace containing the gluon fields $F(0)$ and $F(\xi)$ and also in the additional color traced parts. 

Just as for the pure gauge loops in the previous subsection, the integrated result and the first transverse moment only have gluonic poles and derivatives in the part containing $F(0)$ and $F(\xi)$. Since the additional traced loops reduce to unity after $p_\st$-integration, the color structures are the same as for type 1. For the second moment, there is now one additional structure since the two gluonic poles can also be in the traced loop,
\begin{equation}
\Gamma_{GG,3}^{\alpha_1\alpha_2}\rightarrow \frac{2}{N_c}\tr_c\Big\lgroup\left\{ G_\st^{\alpha_1}(\xi), G_\st^{\alpha_2}(\xi)\right\}\Big\rgroup\tr_c\Big\lgroup F(0)F(\xi)\Big\rgroup . 
\end{equation}
In the triple weighted matrix elements with only gluonic poles, one can have two or three gluonic poles in the traced loop, leading to three new operator structures with gluonic pole rank three,
\begin{subequations}
\begin{align}
\Gamma_{GGG,3}^{\{\alpha_1\alpha_2\alpha_3\}}&\rightarrow 
\frac{2}{N_c}\tr_c\Big\lgroup\big\{ G_\st^{\{\alpha_1}(\xi),G^{\alpha_2}(\xi)\big\}\Big\rgroup
\tr_c\Big\lgroup F(0)\big[ G_\st^{\alpha_3\}}(\xi),F(\xi)\big] \Big\rgroup , \\
\Gamma_{GGG,4}^{\{\alpha_1\alpha_2\alpha_3\}}&\rightarrow 
\frac{2}{N_c}\tr_c\Big\lgroup\big\{ G_\st^{\{\alpha_1}(\xi), G_\st^{\alpha_2}(\xi)\big\}\Big\rgroup
\tr_c\Big\lgroup F(0)\left\{ G_\st^{\alpha_3\}}(\xi),F(\xi)\right\} \Big\rgroup , \\
\Gamma_{GGG,5}^{\alpha_1\alpha_2\alpha_3}&\rightarrow 
\frac{2}{N_c}\tr_c\Big\lgroup\big\{ G_\st^{\alpha_1}(\xi),\big\{ G_\st^{\alpha_2}(\xi), G_\st^{\alpha_3}(\xi)\big\}\big\}\Big\rgroup
\tr_c\Big\lgroup F(0)F(\xi) \Big\rgroup .
\end{align}
\end{subequations}
On top of this, for triple weighting an operator structure including one partial derivative appears in the combination
\begin{eqnarray}
\widetilde\Gamma_{\{\partial GG\},3}^{\{\alpha_1\alpha_2\alpha_3\}}&\rightarrow & 
\tfrac{6}{N_c}\tr_c\Big\lgroup\big\{ G_\st^{\{\alpha_1}(\xi), G_\st^{\alpha_2}(\xi)\big\}\Big\rgroup\tr_c\Big\lgroup F(0)\big[ i\partial_{\st}^{\alpha_3\}},F(\xi) \big]\Big\rgroup \nonumber \\
&&\mbox{} +\tfrac{2}{N_c}
\tr_c\Big\lgroup\big[i\partial_{\st}^{\{\alpha_1},\big\{ G_\st^{\alpha_2}(\xi), G_\st^{\alpha_3\}}(\xi)\big\}\big]\Big\rgroup
\tr_c\Big\lgroup F(0)F(\xi)\Big\rgroup \nonumber \\
&&\mbox{} +\tfrac{2}{N_c}
\tr_c\Big\lgroup\big\{ G_\st^{\{\alpha_1}(\xi),\big[i\partial_{\st}^{\alpha_2}, G_\st^{\alpha_3\}}(\xi)\big]\big\}\Big\rgroup
\tr_c\Big\lgroup F(0)F(\xi)\Big\rgroup .
\end{eqnarray}
This is derived in a straightforward way using derivatives as in Table~\ref{t:operators} but now for a product of two factors, $O_1(\xi) = 1$ and $O_2(\xi) = F(\xi)$, with the appropriate gauge links.

\subsection{Weighting for type 3 correlators}
Taking transverse moments for type 3 correlators can never give the correlators that were found for the type 1 and type 2 correlators, since the gluon fields $F(0)$ and $F(\xi)$ are located in different color traces. When constructing transverse moments, however, we need at least one gluonic pole matrix element (with color octet structure) in each of the traces. The simplest operator structure appearing in double weighting is
\begin{equation}
\Gamma_{GG,4}^{\{\alpha_1\alpha_2\}}\rightarrow \tfrac{2}{N_c}
\tr_c\Big\lgroup F(0)\, G_\st^{\{\alpha_1}(\xi)\Big\rgroup
\tr_c\Big\lgroup\big\{ G_\st^{\alpha_2\}}(\xi),F(\xi)\big\}\Big\rgroup .
\end{equation}
For triple weighting one finds the gluonic pole operator structures
\begin{subequations}
\begin{align}
\Gamma_{GGG,6}^{\{\alpha_1\alpha_2\alpha_3\}}&\rightarrow \tfrac{2}{N_c}
\tr_c\Big\lgroup F(0)\, \big\{ G_\st^{\{\alpha_1}(\xi), G_\st^{\alpha_2}(\xi)\big\}\Big\rgroup
\tr_c\Big\lgroup\big\{ G_\st^{\alpha_3\}}(\xi),F(\xi)\big\}\Big\rgroup, \\
\Gamma_{GGG,7}^{\{\alpha_1\alpha_2\alpha_3\}}&\rightarrow \tfrac{2}{N_c}
\tr_c\Big\lgroup F(0)\, G_\st^{\{\alpha_1}(\xi) \Big\rgroup
\tr_c\Big\lgroup\big\{ G_\st^{\alpha_2}(\xi),\big\{ G_\st^{\alpha_3\}}(\xi),F(\xi)\big\}\big\}\Big\rgroup.
\end{align}
\end{subequations}
On top of this, one will also find the matrix element
\bea
\widetilde\Gamma_{\{\partial GG\},4}^{\{\alpha_1\alpha_2\alpha_3\}}&\rightarrow &\tfrac{3}{N_c}
\tr_c\Big\lgroup F(0)\, \big[i\partial_\st^{\{\alpha_1},G_\st^{\alpha_2}(\xi)\big]\Big\rgroup
\tr_c\Big\lgroup\big\{ G_\st^{\alpha_3\}}(\xi),F(\xi)\big\}\Big\rgroup \nonumber \\
&&+\tfrac{3}{N_c}
\tr_c\Big\lgroup F(0)\, G_\st^{\{\alpha_1}(\xi)\Big\rgroup
\tr_c\Big\lgroup\big\{ G_\st^{\alpha_2}(\xi),\big[i\partial_{\st}^{\alpha_3\}},F(\xi)\big]\big\}\Big\rgroup \nonumber \\
&&+\tfrac{3}{N_c}\tr_c\Big\lgroup F(0) G_\st^{\{\alpha_1}(\xi)\Big\rgroup
\tr_c\Big\lgroup\big[i\partial_{\st}^{\alpha_2},\big\{ G_\st^{\alpha_3\}}(\xi),F(\xi)\big\}\big]\Big\rgroup,
\eea
where we have absorbed a factor of 3 in the definition, again as the natural binomial coefficient appearing when taking derivatives of a product.

\section{Defining TMDs\label{s:definingTMDs}}
In order to define gluon TMD PDFs, we will make an expansion into $p_\st^2$-dependent correlators multiplying symmetric traceless tensors in transverse momentum space, where we include correlators with all possible values of gluonic pole rank or partial derivative rank, the sum of which defines the rank of the universal TMD correlators,
\bea
\Gamma^{[U]}(x,p_\st) &\ =\ &
\Gamma(x,p_\st^2) 
+ \frac{p_{\st i}}{M}\,\widetilde\Gamma_\partial^{i}(x,p_\st^2)
+ \frac{p_{\st ij}}{M^2}\,\widetilde\Gamma_{\partial\partial}^{ij}(x,p_\st^2)
+ \frac{p_{\st ijk}}{M^3}\,\widetilde\Gamma_{\partial\partial\partial}^{\,ijk}(x,p_\st^2) 
+\ldots 
\nonumber \\ &\quad +&
\sum_c C_{G,c}^{[U]}\left\lgroup\frac{p_{\st i}}{M}\,\Gamma_{G,c}^{i}(x,p_\st^2)
+ \frac{p_{\st ij}}{M^2}\,\widetilde\Gamma_{\{\partial G\},c}^{\,ij}(x,p_\st^2)
+ \frac{p_{\st ijk}}{M^3}\,\widetilde\Gamma_{\{\partial\partial G\},c}^{\,ijk}(x,p_\st^2) + \ldots\right\rgroup
\nonumber \\ &\quad +&
\sum_c C_{GG,c}^{[U]}\left\lgroup\frac{p_{\st ij}}{M^2}\,\Gamma_{GG,c}^{ij}(x,p_\st^2)
+ \frac{p_{\st ijk}}{M^3}\,\widetilde\Gamma_{\{\partial GG\},c}^{\,ijk}(x,p_\st^2)+ \ldots\right\rgroup
\nonumber \\ &\quad +&
\sum_c C_{GGG,c}^{[U]}\left\lgroup\frac{p_{\st ijk}}{M^3}\,\Gamma_{GGG,c}^{ijk}(x,p_\st^2)
+ \ldots \right\rgroup + \ldots \, ,
\label{e:TMDstructure}
\eea
where we have suppressed the indices $\mu$ and $\nu$ of the gluon fields for readability. Just as for the transverse moments in the previous section, multiple color structures are possible in Eq.~\ref{e:TMDstructure} for the matrix elements, hence the summation over the color structures $c$. The allowed color structures and the meaning of operator combinations like $\{\partial\partial G\}$ are the same ones that have been discussed in section~\ref{s:colorstructures}. Note that since the tensors $p_\st^{ij}$ and $p_\st^{ijk}$ on the rhs of Eq.~\ref{e:TMDstructure} are traceless and symmetric, the correlators to which they correspond can be defined to be traceless as well (see Appendix~\ref{A:operator} for their definition). This is in fact even necessary in order to make the identification of these correlators in terms of TMDs unique.

Equivalently with traceless symmetric tensors, we can use the real and imaginary part of $\vert p_\st\vert^m e^{im\varphi}$ as the two independent components of the symmetric traceless tensor of rank $m$. Just as was done for quarks in Ref.~\cite{Buffing:2012sz}, an identification should be made between the matrix elements and the gluon TMDs. We already mentioned the rank of the TMD correlators, which is determined by the number of transverse fields $i\partial_\st$ = $iD_\st-A_\st$ and gluonic poles $G$ in the matrix elements and equals the rank of the (symmetric and traceless) tensor constructed from the transverse momenta. Secondly the behavior of both matrix elements and the TMDs under time-reversal symmetry will be used to identify specific distribution functions in the parametrization. A gluonic pole contribution is T-odd. Therefore, all matrix elements containing an odd number of gluonic poles are T-odd, while all matrix elements containing an even number of gluonic poles (or no gluonic poles at all) are T-even. Using this identification, it is possible to determine for each TMD separately in which box(es) in Table~\ref{t:spinhalfcol-1} it should belong, the result of which can be seen in Table~\ref{t:gluonPDF-1}.

By comparing the Eqs.~\ref{e:GluonCorr} and \ref{e:TMDstructure} it becomes clear that for gluons contributions up to rank 3 have to be taken into account. The rank 0 contribution is given by
\bea
2x\,\Gamma^{\mu\nu [U]}(x{,}p_\st) &=& -g_T^{\mu\nu}\,f_1^{g [U]}(x{,}p_\st^2)+i\epsilon_T^{\mu\nu}\;S_{\sL} g_{1L}^{g [U]}(x{,}p_\st^2).
\label{e:rank0}
\eea
In order to analyze the rank 1 contributions, weighting of the correlator $\Gamma^{[U]}(x,p_\st)$ with one factor of $p_T$ is required, yielding an expression of the form
\bea
\frac{p_\st^\alpha}{M}\,\Gamma^{[U]}(x,p_\st) &\ =\ &
\frac{p_{\st}^\alpha}{M}\,\Gamma(x,p_\st^2)
-\widetilde\Gamma_{\partial}^{\alpha (1)}(x,p_\st^2)
- C_{G}^{[U]}\,\Gamma_G^{\alpha (1)}(x,p_\st^2)
\nonumber\\[3pt] &\quad \mbox{} +&
\frac{p^{\ \alpha}_{\st \,i}}{M^2}
\,\widetilde \Gamma_\partial^i(x,p_\st^2)
+C_{G}^{[U]}\,\frac{p^{\ \alpha}_{\st \,i}}{M^2}
\, \Gamma_G^i(x,p_\st^2) + \ldots \, ,
\label{e:fullfirst}
\eea
where the dots stand for contributions of rank 2 and higher that do not survive the $p_T$-integration. The second and third term in the first line contain the lowest rank transverse moments. Using the behavior under time-reversal of these matrix elements, one finds the rank 1 contributions
\bea
&&2x\,\frac{p_{\st i}}{M}\,\widetilde\Gamma_\partial^{i\,\mu\nu}(x,p_\st^2) = -i\epsilon_T^{\mu\nu}\frac{p_{\st}\cdot S_{\st}}{M}\;g_{1T}^{g}(x{,}p_\st^2), \\
&&2x\,\frac{p_{\st i}}{M}\,\Gamma_{G,c}^{i\,\mu\nu}(x,p_\st^2) =g_T^{\mu\nu}\frac{\epsilon_T^{p_TS_T}}{M}\, f_{1T}^{\perp g(Ac)}(x{,}p_\st^2)-\frac{\epsilon_T^{p_T\{\mu}S_T^{\nu\}}{+}\epsilon_T^{S_T\{\mu}p_T^{\nu\}}}{4M}\; h_{1T}^{g(Ac)}(x{,}p_\st^2).
\label{e:rank1}
\eea
Without going into the technical details, similar calculations for rank 2 and rank 3 contributions yield
\bea
&&2x\,\frac{p_{\st ij}}{M^2}\,\widetilde\Gamma_{\partial\partial}^{ij\,\mu\nu}(x,p_\st^2) = p_{\st ij}g_T^{i\mu}g_T^{j\nu}\;h_1^{\perp g (A)}(x{,}p_\st^2), \\
&&2x\,\frac{p_{\st ij}}{M^2}\,\widetilde\Gamma_{\{\partial G\},c}^{ij\,\mu\nu}(x,p_\st^2) = -\frac{p_{\st ij}\epsilon_T^{i\{\mu}g_T^{\nu\}j}S_{\sL}}{2M^2}\; h_{1L}^{\perp g (Ac)}(x{,}p_\st^2), \label{e:PhipartialG} \\
&&2x\,\frac{p_{\st ij}}{M^2}\,\Gamma_{GG,c}^{ij\,\mu\nu}(x,p_\st^2) = p_{\st ij}g_T^{i\mu}g_T^{j\nu}\; h_1^{\perp g (Bc)}(x{,}p_\st^2), \\
&&2x\,\frac{p_{\st ijk}}{M^3}\,\widetilde\Gamma_{\partial\partial\partial}^{ijk\,\mu\nu}(x,p_\st^2) = 0, \\
&&2x\,\frac{p_{\st ijk}}{M^3}\,\widetilde\Gamma_{\{\partial\partial G\},c}^{ijk\,\mu\nu}(x,p_\st^2) = p_{\st ijk}\frac{\epsilon_T^{i\{\mu}g_T^{\nu\}j}}{2M^2}\frac{S_{\st}^{k}}{M}\; h_{1T}^{\perp g(Ac)}(x{,}p_\st^2), \\
&&2x\,\frac{p_{\st ijk}}{M^3}\,\widetilde\Gamma_{\{\partial GG\},c}^{ijk\,\mu\nu}(x,p_\st^2) = 0, \\
&&2x\,\frac{p_{\st ijk}}{M^3}\,\Gamma_{GGG,c}^{ijk\,\mu\nu}(x,p_\st^2) = p_{\st ijk}\frac{\epsilon_T^{i\{\mu}g_T^{\nu\}j}}{2M^2}\frac{S_{\st}^{k}}{M}\; h_{1T}^{\perp g(Bc)}(x{,}p_\st^2),
\eea
where we have used Eq.~\ref{e:GluonCorr} for the gluon parametrization. The labels $A$ and $B$, with or without an additional color index $c$, are used to label the different matrix elements and the multiple color possibilities for TMDs with the same rank, as discussed in the previous section. There are two color possibilities for $f_{1T}^{\perp g(Ac)}$, $h_{1T}^{g(Ac)}$, $h_{1L}^{\perp g(Ac)}$ and $h_{1T}^{\perp g(Ac)}$, four for $h_1^{\perp g (Bc)}$ and seven for $h_{1T}^{\perp g(Bc)}$.

We can also look at each structure in the matrix element separately. In this paper nonuniversal TMDs were introduced, namely $f_{1T}^{\perp g[U]}$, $h_{1T}^{g[U]}$, $h_{1L}^{\perp g[U]}$, $h_1^{\perp g[U]}$ and $h_{1T}^{\perp g[U]}$. Using the complete list of structures for the type 1, type 2 and type 3 correlators, it is possible to give the expressions of these nonuniversal TMDs in terms of the universal definite rank TMDs,
\bea
f_{1T}^{\perp g[U]}(x,p_\st^2)&=&\sum_{c=1}^2 C_{G,c}^{[U]}\,f_{1T}^{\perp g(Ac)}(x,p_\st^2), \\
h_{1T}^{g[U]}(x,p_\st^2)&=&\sum_{c=1}^2 C_{G,c}^{[U]}\,h_{1T}^{g(Ac)}(x,p_\st^2), \\
h_{1L}^{\perp g[U]}(x,p_\st^2)&=&\sum_{c=1}^2 C_{G,c}^{[U]}\,h_{1L}^{\perp g(Ac)}(x,p_\st^2), \\
h_1^{\perp g[U]}(x,p_\st^2)&=&h_1^{\perp g (A)}(x,p_\st^2)+\sum_{c=1}^{4}C_{GG,c}^{[U]}\,h_1^{\perp g (Bc)}(x,p_\st^2), \\
h_{1T}^{\perp g[U]}(x,p_\st^2)&=&\sum_{c=1}^2 C_{G,c}^{[U]}\,h_{1T}^{\perp g(Ac)}(x,p_\st^2)+\sum_{c=1}^{7}C_{GGG,c}^{[U]}\,h_{1T}^{\perp g(Bc)}(x,p_\st^2).
\eea

A similar expansion as in Eq.~\ref{e:TMDstructure} can also be made for the fragmentation correlator $\Delta^{g}(z,k_\st)$ as for instance given in Ref.~\cite{Mulders:2000sh}. For fragmentation functions, the gluonic pole matrix elements vanish~\cite{Gamberg:2010gp,Meissner:2008yf,Gamberg:2008yt,Collins:2004nx,Metz:2002iz}, hence there is no longer any process dependence. As a result, all fragmentation TMDs are universal. The assignment of gluon TMD PFFs can be seen in Table~\ref{t:gluonFF-1}. Although gluonic pole matrix elements vanish, there are T-odd functions that appear in the parametrization of the $\widetilde\Delta_{\partial\ldots\partial}(z,k_\st^2)$ matrix elements. 

\begin{table}[!tb]
\centering
\begin{tabular}{|m{10mm}|m{23mm}|m{23mm}|m{23mm}|m{23mm}|m{23mm}|}
\hline
& \multicolumn{4}{|c|}{RANK} \\ \hline
\# GPs &\qquad\quad 0 & \qquad\quad 1 & \qquad\quad 2 & \qquad\quad 3 
\\ \hline
0 
&$\Gamma(x,p_\st^2)$
&$\widetilde\Gamma_\partial$
&$\widetilde\Gamma_{\partial\partial}$
&$\widetilde\Gamma_{\partial\partial\partial}$
\\[2pt]
\hline
1
& &$C_{G,c}^{[U]}\Gamma_{G,c}$
&$C_{G,c}^{[U]}\widetilde\Gamma_{\{\partial G\},c}$
&$C_{G,c}^{[U]}\widetilde\Gamma_{\{\partial\partial G\},c}$
\\[2pt]
\hline
2
& & &$C_{GG,c}^{[U]}\Gamma_{GG,c}$
&$C_{GG,c}^{[U]}\widetilde\Gamma_{\{\partial GG\},c}$
\\[2pt]
\hline
3
& & & &$C_{GGG,c}^{[U]}\Gamma_{GGG,c}$
\\[2pt] 
\hline
\end{tabular}
\parbox{0.85\textwidth}{
\caption{
The matrix elements for gluon TMD PDFS, ordered by gluonic pole rank and number of transverse weightings. The explicit expansion of all correlators in universal correlators multiplied with gluonic pole factors is given in the text. Note that the gluonic pole coefficients are equal for correlators in the same row.
\label{t:spinhalfcol-1}}}
\end{table}

\begin{table}[!tb]
\centering
\begin{tabular}{|m{10mm}|p{23mm}|p{23mm}|p{23mm}|p{23mm}|}
\hline
& \multicolumn{4}{|c|}{RANK OF TMD PDFs FOR GLUONS} \\ \hline
\# GPs &\qquad\quad 0 & \qquad\quad 1 & \qquad\quad 2 & \qquad\quad 3 
\\ \hline
0 
&$f_1^g$, $g_{1}^g$
&$g_{1T}^g$
&$h_1^{\perp g(A)}$
&
\\[2pt]
\hline
1
&&$f_{1T}^{\perp g(Ac)}$, $h_{1T}^{g(Ac)}$
&$h_{1L}^{\perp g(Ac)}$
&$h_{1T}^{\perp g(Ac)}$
\\[2pt]
\hline
2
&&&$h_1^{\perp g(Bc)}$
&
\\[2pt]
\hline
3
&&&&$h_{1T}^{\perp g(Bc)}$
\\[2pt]\hline
\end{tabular}
\parbox{0.85\textwidth}{
\caption{
The operator assignments of TMD PDFs for gluons. The index $c$ for some labels \textit{A} and \textit{B} indicate that there are multiple contributions for that TMD PDF of that rank due to the presence of multiple color structures.
\label{t:gluonPDF-1}}}
\end{table}

\begin{table}[!tb]
\centering
\begin{tabular}{|p{10mm}|p{23mm}|p{23mm}|p{23mm}|p{23mm}|}
\hline
&\multicolumn{4}{|c|}{RANK OF TMD PFFs FOR GLUONS} \\ 
\hline
\# GPs &\qquad\quad 0 & \qquad\quad 1 & \qquad\quad 2 & \qquad\quad 3 
\\ \hline
0
&$D_1^g$, $G_{1}^g$
&$D_{1T}^{\perp g}$, $G_{1T}^g$, $H_{1T}^g$
&$H_1^{\perp g}$, $H_{1L}^{\perp g}$
&$H_{1T}^{\perp g}$
\\[2pt]
\hline
\end{tabular}
\parbox{0.85\textwidth}{
\caption{
The matrix element assignment for gluon fragmentation functions. All gluonic pole matrix elements vanish.
\label{t:gluonFF-1}}}
\end{table}

\begin{figure}
\epsfig{file=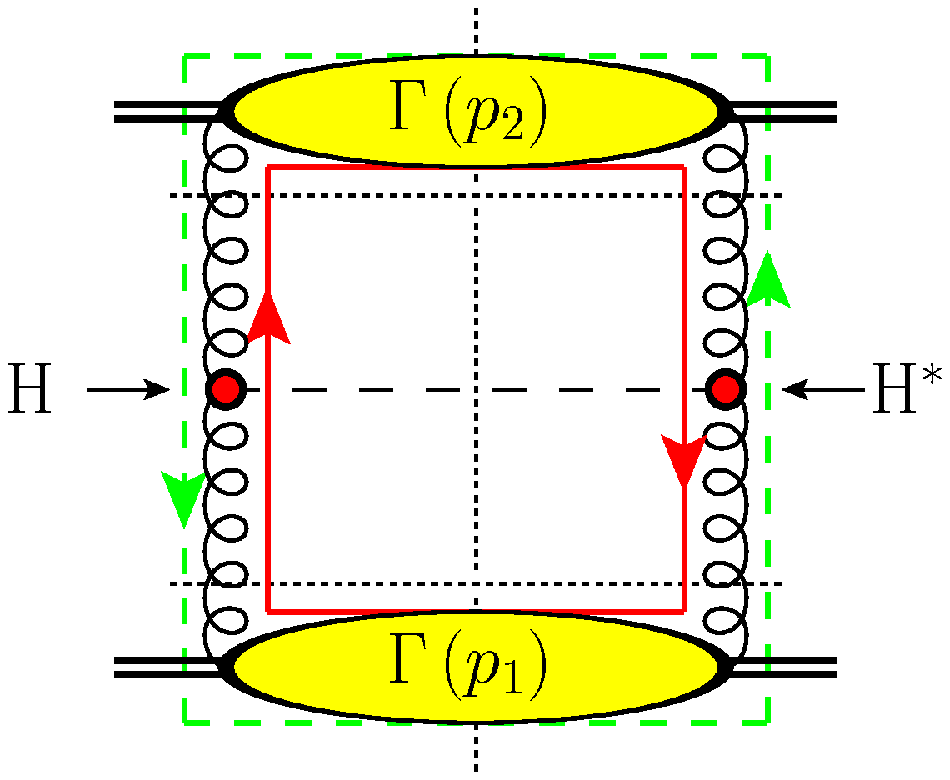,width=0.3\textwidth}
\hspace{15mm}
\epsfig{file=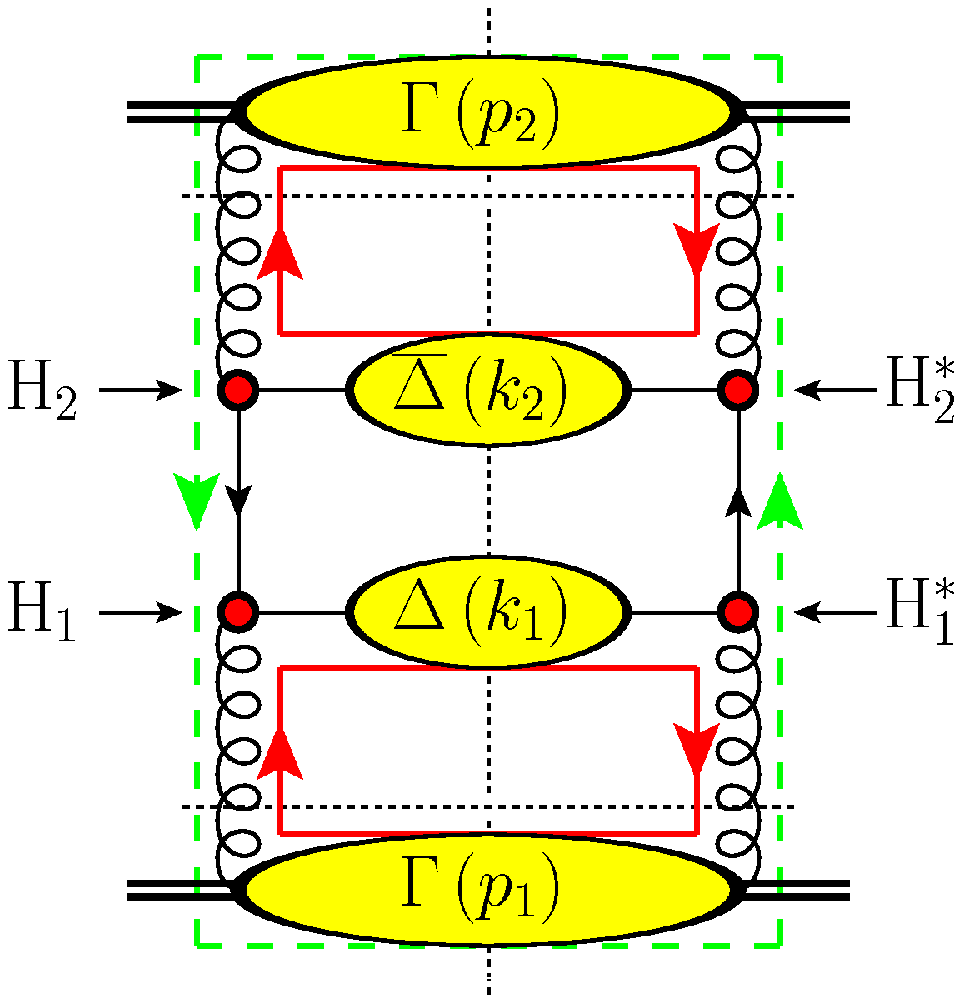,width=0.3\textwidth}
\\[1mm]
(a)\hspace{66mm} (b)
\caption{\label{f:colorflow}
A number of color flow structures illustrated with red continuous and green dashed lines. Spectator colors have not been indicated. In (a) the color flow for a $gg\rightarrow \textit{colorless final state}$ process (e.g. $gg\rightarrow \text{Higgs}$) can be seen, which comes with the $U^{[-,-^{\dagger}]}$ gauge link. In (b) the dominant color flow for a particular $gg\rightarrow q\overline{q}$ diagram is illustrated. There are three more contributing color flow structures, but they are suppressed by one or more factors of $N_c$.}
\end{figure}

Knowing how to expand the correlators $\Gamma^{[U]}(x,p_\st)$ into universal definite rank TMD correlators $\Gamma_{O}^{i_1\ldots i_m}(x,p_\st^2)$, we would like to invert these relations to obtain the actual operator expressions of these definite rank correlators in terms of TMD correlators with particular gauge links. Knowing the links one can identify high-energy processes in which to measure them or one may (eventually) be able to do a lattice calculation of these correlators. As explained in many other papers, it is the color structure of the hard amplitude that determines the gauge link structure, illustrated for two particular diagrams in Fig.~\ref{f:colorflow}.

For the proper identification of the TMD correlators and the TMD PDFs depending on $x$ and $p_\st^2$ with operator matrix elements, we actually need the results after doing only the azimuthal integration,
\be 
\int \frac{d\varphi}{2\pi}\ \frac{p_\st^{\alpha_1\ldots\alpha_m}}{M^m}\,\frac{p_{\st\,i_1\ldots i_m}}{M^m}\,\Gamma_{\dots}^{i_1\ldots i_m}(x,p_\st^2)
\ = \ \left(\frac{p_\st^2}{2M^2}\right)^m\,\Gamma_{\ldots}^{\alpha_1\ldots\alpha_m}(x,p_\st^2)
\ = \ (-)^m\,\Gamma_{\dots}^{\alpha_1\ldots\alpha_m\,(m)}(x,p_\st^2),
\label{e:tensormult}
\ee
with the transverse moments on the rhs defined as in Eq.~\ref{e:transversemoments} and where the subscript is the appropriate combination of $\partial$ and $G$ operators. To get the integrated transverse moments one needs an additional integration $\int \pi \,d\vert p_\st\vert^2$. To perform the $\varphi$-integration, we use the fact that the independent components of $p_\st^{\alpha_1\ldots \alpha_m}$ can for $m\ge 1$ be written as
\be 
p_\st^{\alpha_1\ldots \alpha_m} \quad \Longleftrightarrow\frac{\vert p_\st\vert^m}{2^{m-1}}\,e^{\pm im\varphi}.
\ee
As an example, we obtain from Eq.~\ref{e:fullfirst} with the help of the coefficients in the Tables~\ref{t:gpfactors1} and \ref{t:gpfactors2}, the relations
\bea
&&
\frac{p_\st^2}{2M^2}\,\widetilde\Gamma_{\partial}^{\alpha}(x,p_\st^2) 
= \frac{1}{2}\int\frac{d\varphi}{2\pi}\ \frac{p_\st^\alpha}{M}\left\lgroup \Gamma^{[+,+^\dagger]}(x,p_\st) + \Gamma^{[-,-^\dagger]}(x,p_\st)\right\rgroup ,
\label{e:linking-1}
\\&&
\frac{p_\st^2}{2M^2}\,\Gamma_{G,1}^{\alpha}(x,p_\st^2) 
= \frac{1}{2}\int\frac{d\varphi}{2\pi}\ \frac{p_\st^\alpha}{M}\left\lgroup \Gamma^{[+,+^\dagger]}(x,p_\st) - \Gamma^{[-,-^\dagger]}(x,p_\st)\right\rgroup .
\label{e:linking-2}
\eea
Note, however, that these expressions are not unique. For instance in Eq.~\ref{e:linking-1} one could have used a correlator with an additional Wilson loop, since one finds from Tables~\ref{t:gpfactors1} and \ref{t:gpfactors2} that there are relations like
\be
\int\frac{d\varphi}{2\pi}\ \frac{p_\st^\alpha}{M}\left\lgroup \Gamma^{[+,+^\dagger(\square)]} - \Gamma^{[+,+^\dagger]}\right\rgroup = 0.
\ee
The actual $\Gamma^{[+,+^\dagger (\square)]}(x,p_\st)$ has a different azimuthal dependence as compared to $\Gamma^{[+,+^\dagger]}(x,p_\st)$, but in the $\varphi$-integrated situation they yield the same results for the first harmonic dependence in $\varphi$. The correlators differ in the second and third harmonics. We can make this even slightly more explicit. Looking at the operator structure as discussed in Section~\ref{s:colorstructures}, we note that the first moment of the TMD correlator with an additional Wilson loop contains the operator combination
\be
\tr_c\Big\lgroup F(0)U_{[0,\xi]}^{[+]}F(\xi)U_{[\xi,0]}^{[+]}\Big\rgroup\frac{1}{N_c}\tr_c\Big\lgroup U_{[0,\xi]}^{[+]} G_\st^{\alpha}(\xi)U_{[\xi,0]}^{[-]}\Big\rgroup, \label{e:++dag(loop)_1}
\ee
of which the matrix element is nonzero, becoming zero after azimuthal averaging. Terms of this type also were considered as {\it junk-TMD} in Ref.~\cite{Bomhof:2007xt}. We note, however, that in the present treatment all such contributions are contained in the complete expansion in Eq.~\ref{e:TMDstructure}. To find a precise form for the difference for the second and third harmonic dependence, one has to use the appropriate gluonic pole factors that are given in Section~\ref{s:colorstructures} and write down the first moment completing Eq.~\ref{e:fullfirst}. This unintegrated first moment not only contains the rank 1 functions, but also functions of other rank. Only after $\varphi$-integration, one is left with functions of one particular rank.

\section{Conclusions}
In this paper we have introduced gluon TMDs of definite rank $m$. These appear in the expansion of an arbitrary gauge link dependent correlator as gauge link independent functions multiplied with irreducible traceless tensors $p_\st^{i_1\ldots i_m}$. The expansion contains the full set of matrix elements appearing in the description of gluon transverse momentum dependent parton distribution functions (TMD PDFs). These matrix elements can be classified by the rank of the operator combination it contains. This rank is equal to the sum of the number of gluonic pole operators and color gauge-invariant partial derivative operators. Using Lorentz invariance, hermiticity, inversion and time-reversal symmetry, their matrix elements can be identified with TMDs. Like for the quark TMDs, multiple color structures could appear for matrix elements containing a certain number of gluonic poles, which increases the number of (independently) contributing process independent leading twist gluon TMDs from 8 to 23. Nevertheless, the introduction of universal definite rank TMDs improves on the situation in which one has eight types of gauge link dependent TMDs, $f_{\ldots}^{\ldots g\,[U]}(x,p_\st)$. Although we get 23 functions $f_{\ldots}^{\ldots g\,(\ldots)}(x,p_\st^2)$ carrying additional indices like $(Ac)$, they are universal. In particular, the gauge link dependence is contained in calculable gluonic pole factors, which can be obtained for any hard process.

In classifying contributions using their rank, we find two rank 0 functions, the $f_1^g$ and $g_{1L}^{\perp g}$, both of which come without process dependent factor. We find three types of rank 1 functions. One of them, the $g_{1T}^g$ TMD, can be identified with the matrix element $\widetilde{\Gamma}_{\partial}$ and does not have a gluonic pole factor. The TMDs $f_{1T}^{g\,(Ac)}$ and $h_{1T}^{g\,(Ac)}$ correspond to the matrix elements $\Gamma_{G,c}$, where $c$ labels the two ways to neutralize three gluon fields using SU(3) structure constants $f$ or $d$. The emerging of these two TMDs with different color structures has been introduced earlier in Ref.~\cite{Bomhof:2007xt}. For rank 2, we find two functions, namely $h_{1}^{\perp g\,(\ldots)}$ and $h_{1L}^{\perp g\,(\ldots)}$. The $h_{1}^{\perp g\,(Ac)}$ and $h_{1}^{\perp g\,(Bc)}$ TMDs correspond to the matrix elements $\widetilde{\Gamma}_{\partial\partial}$ and $\Gamma_{GG,c}$, respectively, the latter having four different color structures. The $h_{1L}^{\perp g\,(Ac)}$ functions can be identified with the matrix elements $\widetilde{\Gamma}_{\{\partial G\},c}$. Just as for the other TMDs that can be identified with matrix elements containing just one gluonic pole, there are only two color structures for this TMD. For rank 3, the dependence of the TMD correlator on transverse momentum and spin allows just a single structure. Nevertheless it gives rise to nine universal, T-odd functions denoted $h_{1T}^{\perp g\,(Ac)}$ and $h_{1T}^{\perp g\,(Bc)}$. These correspond to the two matrix elements in $\widetilde{\Gamma}_{\{\partial\partial G\},c}$ and the seven matrix elements in $\Gamma_{GGG,c}$.

This new expansion in universal TMDs is of use to any experiment or analysis studying high-energy processes with the aim of extracting TMDs and comparing them to the results of extractions of TMDs using a different process, since it is possible to calculate which combinations of TMDs are relevant for a particular high-energy process. These will be given in a forthcoming publication. We do caution, however, that the color dependent factors multiplying the universal functions as defined in this paper are a tree level result. They are the lowest order coefficients in a full calculation. The relevant parts that have been resummed are the collinear gluons needed to obtain the appropriate color gauge invariant matrix elements in the TMDs.

\section*{Acknowledgements}
This research is part of the research program of the ``Stichting voor Fundamenteel Onderzoek der Materie (FOM)", which is financially supported by the ``Nederlandse Organisatie voor Wetenschappelijk Onderzoek (NWO)". We also acknowledge support of the FP7 EU-programme HadronPhysics3 (contract no 283286) and QWORK (contract 320389). AM thanks the Alexander von Humboldt Fellowship for Experienced Researchers, Germany, for support. We acknowledge discussions with Wilco den Dunnen, Cedric Lorc\'{e} and Andreas Metz. All figures were made using JaxoDraw~\cite{Binosi:2003yf,Binosi:2008ig}.

\appendix

\section{Operator combinations\label{A:operator}}
In this appendix, we aim to give a list of relations for all the correlators that we have defined in section~\ref{s:colorstructures} in terms of pairs of gauge link structures that are a time-reversal couple, which e.g. for the rank 0 correlator on the rhs of Eq.~\ref{e:TMDstructure} is given by
\bea
\Gamma(x,p_\st^2) &=& \frac{1}{2}\int\frac{d\varphi}{2\pi}\left\lgroup \Gamma^{[+,+^\dagger]}(x,p_\st)+\Gamma^{[-,-^\dagger]}(x,p_\st)\right\rgroup . \label{e:cor}
\eea
The correlators are symmetric and traceless by definition, which implies e.g.
\be
\widetilde\Gamma_{\partial\partial}^{\alpha_1\alpha_2}(x,p_\st^2)=\frac{1}{2}\,\widetilde\Gamma_{\partial\partial}^{\{\alpha_1\alpha_2\}}(x,p_\st^2)-\frac{1}{2}g_\st^{\alpha_1\alpha_2}\widetilde\Gamma_{\partial\partial}^{ii}(x,p_\st^2)
\ee
for correlators of rank 2. A similar expression holds for rank 3 contributions, containing a symmetrization over three indices and multiple trace terms. In this appendix we will furthermore use the symmetric traceless tensors of rank 2 and 3 given explicitly by
\bea
p_\st^{ij} &=& p_\st^i p_\st^j - \frac{1}{2}\,p_\st^2\,g_\st^{ij}, \\
p_\st^{ijk} &=& p_\st^i p_\st^j p_\st^k - \frac{1}{4}\,p_\st^2\left(g_\st^{ij}p_\st^k + g_\st^{ik}p_\st^j + g_\st^{jk}p_\st^i\right).
\eea

\subsection{Rank 1 contributions}
The correlators for rank 1 contributions are given by
\bea
\widetilde\Gamma_{\partial}^{\alpha_1 (1)}(x,p_\st^2) &=& -\frac{1}{2}\int\frac{d\varphi}{2\pi}\ \frac{p_\st^{\alpha_1}}{M}\left\lgroup \Gamma^{[+,+^\dagger]}(x,p_\st)+\Gamma^{[-,-^\dagger]}(x,p_\st)\right\rgroup, \label{e:cor-d} \\
\Gamma_{G,1}^{\alpha_1 (1)}(x,p_\st^2) &=& -\frac{1}{2}\int\frac{d\varphi}{2\pi}\ \frac{p_\st^{\alpha_1}}{M}\left\lgroup \Gamma^{[+,+^\dagger]}(x,p_\st)-\Gamma^{[-,-^\dagger]}(x,p_\st)\right\rgroup, \label{e:cor-G1} \\
\Gamma_{G,2}^{\alpha_1 (1)}(x,p_\st^2) &=& -\frac{1}{2}\int\frac{d\varphi}{2\pi}\ \frac{p_\st^{\alpha_1}}{M}\left\lgroup \Gamma^{[+,-^\dagger]}(x,p_\st)-\Gamma^{[-,+^\dagger]}(x,p_\st)\right\rgroup, \label{e:cor-G2}
\eea
where the expressions in the Eqs.~\ref{e:cor-G1} and \ref{e:cor-G2} are the same ones that we give in the Eqs.~\ref{e:linking-1} and \ref{e:linking-2}. The $(1)$ in the superscript of the correlators represents a single transverse moment, in general defined as in Eq.~\ref{e:tensormult}. Since tensor multiplications give factors of $\left(p_\st^2/2M^2\right)^m$, as can be seen explicitly in Eq.~\ref{e:tensormult}, an additional factor of $\left(-\right)^m$ has been included for the correlators. This explains the additional minus sign in expressions for the rank 1 and rank 3 correlators.

\subsection{Rank 2 contributions}
For rank 2 contributions, the T-even parts are given by
\bea
\widetilde\Gamma_{\partial\partial}^{\alpha_1\alpha_2(2)}(x,p_\st^2) &=& \frac{9}{16}\int\frac{d\varphi}{2\pi}\ \frac{p_\st^{\alpha_1\alpha_2}}{M^2}\left\lgroup \Gamma^{[+,+^\dagger]}(x,p_\st)+\Gamma^{[-,-^\dagger]}(x,p_\st)\right\rgroup \nonumber \\
&&-\frac{1}{16}\int\frac{d\varphi}{2\pi}\ \frac{p_\st^{\alpha_1\alpha_2}}{M^2}\left\lgroup \Gamma^{[+\Box,+^\dagger\Box^\dagger]}(x,p_\st)+\Gamma^{[-\Box^\dagger,-^\dagger\Box]}(x,p_\st)\right\rgroup, \label{e:cor-dd} \\
\Gamma_{GG,1}^{\alpha_1\alpha_2(2)}(x,p_\st^2) &=& \frac{1}{2}\int\frac{d\varphi}{2\pi}\ \frac{p_\st^{\alpha_1\alpha_2}}{M^2}\left\lgroup \Gamma^{[+,+^\dagger]}(x,p_\st)+\Gamma^{[-,-^\dagger]}(x,p_\st)\right\rgroup - \widetilde\Gamma_{\partial\partial}^{\alpha_1\alpha_2(2)}(x,p_\st^2), \label{e:cor-GG1} \\
\Gamma_{GG,2}^{\alpha_1\alpha_2(2)}(x,p_\st^2) &=& \frac{1}{2}\int\frac{d\varphi}{2\pi}\ \frac{p_\st^{\alpha_1\alpha_2}}{M^2}\left\lgroup \Gamma^{[+,-^\dagger]}(x,p_\st)+\Gamma^{[-,+^\dagger]}(x,p_\st)\right\rgroup - \widetilde\Gamma_{\partial\partial}^{\alpha_1\alpha_2(2)}(x,p_\st^2), \label{e:cor-GG2} \\
\Gamma_{GG,3}^{\alpha_1\alpha_2(2)}(x,p_\st^2) &=& \frac{1}{2}\int\frac{d\varphi}{2\pi}\ \frac{p_\st^{\alpha_1\alpha_2}}{M^2}\left\lgroup \Gamma^{[+,+^\dagger(\Box)]}(x,p_\st)+\Gamma^{[-,-^\dagger(\Box^\dagger)]}(x,p_\st)\right\rgroup \nonumber \\
&&-\widetilde\Gamma_{\partial\partial}^{\alpha_1\alpha_2(2)}(x,p_\st^2) -\Gamma_{GG,1}^{\alpha_1\alpha_2(2)}(x,p_\st^2), \label{e:cor-GG3} \\
\Gamma_{GG,4}^{\alpha_1\alpha_2(2)}(x,p_\st^2) &=& -\frac{1}{4}\int\frac{d\varphi}{2\pi}\ \frac{p_\st^{\alpha_1\alpha_2}}{M^2}\left\lgroup \Gamma^{[(F(\xi)\Box),(F(0)\Box^\dagger)]}(x,p_\st)+\Gamma^{[(F(\xi)\Box^\dagger),(F(0)\Box)]}(x,p_\st)\right\rgroup, \label{e:cor-GG4}
\eea
while the T-odd parts are given by
\bea
\widetilde\Gamma_{\{\partial G\},1}^{\alpha_1\alpha_2(2)}(x,p_\st^2) &=& \frac{1}{2}\int\frac{d\varphi}{2\pi}\ \frac{p_\st^{\alpha_1\alpha_2}}{M^2}\left\lgroup \Gamma^{[+,+^\dagger]}(x,p_\st)-\Gamma^{[-,-^\dagger]}(x,p_\st)\right\rgroup, \label{e:cor-dG1} \\
\widetilde\Gamma_{\{\partial G\},2}^{\alpha_1\alpha_2(2)}(x,p_\st^2) &=& \frac{1}{2}\int\frac{d\varphi}{2\pi}\ \frac{p_\st^{\alpha_1\alpha_2}}{M^2}\left\lgroup \Gamma^{[+,-^\dagger]}(x,p_\st)-\Gamma^{[-,+^\dagger]}(x,p_\st)\right\rgroup. \label{e:cor-dG2}
\eea

\subsection{Rank 3 contributions}
For the rank 3 correlators the T-even contributions can be defined as
\bea
\widetilde\Gamma_{\partial\partial\partial}^{\alpha_1\alpha_2\alpha_3(3)}(x,p_\st^2) &=& -\frac{9}{16}\int\frac{d\varphi}{2\pi}\ \frac{p_\st^{\alpha_1\alpha_2\alpha_3}}{M^3}\left\lgroup \Gamma^{[+,+^\dagger]}(x,p_\st)+\Gamma^{[-,-^\dagger]}(x,p_\st)\right\rgroup \nonumber \\
&&+\frac{1}{16}\int\frac{d\varphi}{2\pi}\ \frac{p_\st^{\alpha_1\alpha_2\alpha_3}}{M^3}\left\lgroup \Gamma^{[+\Box,+^\dagger\Box^\dagger]}(x,p_\st)+\Gamma^{[-\Box^\dagger,-^\dagger\Box]}(x,p_\st)\right\rgroup =0, \label{e:cor-ddd} \\
\widetilde\Gamma_{\{\partial GG\},1}^{\alpha_1\alpha_2\alpha_3(3)}(x,p_\st^2) &=& -\frac{1}{2}\int\frac{d\varphi}{2\pi}\ \frac{p_\st^{\alpha_1\alpha_2\alpha_3}}{M^3}\left\lgroup \Gamma^{[+,+^\dagger]}(x,p_\st)+\Gamma^{[-,-^\dagger]}(x,p_\st)\right\rgroup + \widetilde\Gamma_{\partial\partial\partial}^{\alpha_1\alpha_2\alpha_3(3)}(x,p_\st^2)=0, \label{e:cor-dGG1} \\
\widetilde\Gamma_{\{\partial GG\},2}^{\alpha_1\alpha_2\alpha_3(3)}(x,p_\st^2) &=& -\frac{1}{2}\int\frac{d\varphi}{2\pi}\ \frac{p_\st^{\alpha_1\alpha_2\alpha_3}}{M^3}\left\lgroup \Gamma^{[+,-^\dagger]}(x,p_\st)+\Gamma^{[-,+^\dagger]}(x,p_\st)\right\rgroup + \widetilde\Gamma_{\partial\partial\partial}^{\alpha_1\alpha_2\alpha_3(3)}(x,p_\st^2)=0, \label{e:cor-dGG2} \\
\widetilde\Gamma_{\{\partial GG\},3}^{\alpha_1\alpha_2\alpha_3(3)}(x,p_\st^2) &=& -\frac{1}{2}\int\frac{d\varphi}{2\pi}\ \frac{p_\st^{\alpha_1\alpha_2\alpha_3}}{M^3}\left\lgroup \Gamma^{[+,+^\dagger(\Box)]}(x,p_\st)+\Gamma^{[-,-^\dagger(\Box^\dagger)]}(x,p_\st)\right\rgroup \nonumber \\
&&+\widetilde\Gamma_{\partial\partial\partial}^{\alpha_1\alpha_2\alpha_3(3)}(x,p_\st^2) + \widetilde\Gamma_{\{\partial GG\},1}^{\alpha_1\alpha_2\alpha_3(3)}(x,p_\st^2)=0, \label{e:cor-dGG3} \\
\widetilde\Gamma_{\{\partial GG\},4}^{\alpha_1\alpha_2\alpha_3(3)}(x,p_\st^2) &=& \frac{1}{4}\int\frac{d\varphi}{2\pi}\ \frac{p_\st^{\alpha_1\alpha_2\alpha_3}}{M^3}\left\lgroup \Gamma^{[(F(\xi)\Box),(F(0)\Box^\dagger)]}(x,p_\st)+\Gamma^{[(F(\xi)\Box^\dagger),(F(0)\Box)]}(x,p_\st)\right\rgroup =0. \label{e:cor-dGG4}
\eea
The T-odd pieces can be defined through
\bea
\widetilde\Gamma_{\{\partial\partial G\},1}^{\alpha_1\alpha_2\alpha_3(3)}(x,p_\st^2) &=& -\frac{9}{16}\int\frac{d\varphi}{2\pi}\ \frac{p_\st^{\alpha_1\alpha_2\alpha_3}}{M^3}\left\lgroup \Gamma^{[+,+^\dagger]}(x,p_\st)-\Gamma^{[-,-^\dagger]}(x,p_\st)\right\rgroup \nonumber \\
&&+\frac{1}{48}\int\frac{d\varphi}{2\pi}\ \frac{p_\st^{\alpha_1\alpha_2\alpha_3}}{M^3}\left\lgroup \Gamma^{[+\Box,+^\dagger\Box^\dagger]}(x,p_\st)-\Gamma^{[-\Box^\dagger,-^\dagger\Box]}(x,p_\st)\right\rgroup, \label{e:cor-ddG1} \\
\widetilde\Gamma_{\{\partial\partial G\},2}^{\alpha_1\alpha_2\alpha_3(3)}(x,p_\st^2) &=& -\frac{9}{16}\int\frac{d\varphi}{2\pi}\ \frac{p_\st^{\alpha_1\alpha_2\alpha_3}}{M^3}\left\lgroup \Gamma^{[+,-^\dagger]}(x,p_\st)-\Gamma^{[-,+^\dagger]}(x,p_\st)\right\rgroup \nonumber \\
&&+\frac{1}{48}\int\frac{d\varphi}{2\pi}\ \frac{p_\st^{\alpha_1\alpha_2\alpha_3}}{M^3}\left\lgroup \Gamma^{[+\Box,-^\dagger\Box]}(x,p_\st)-\Gamma^{[-\Box^\dagger,+^\dagger\Box^\dagger]}(x,p_\st)\right\rgroup, \label{e:cor-ddG2} \\
\Gamma_{GGG,1}^{\alpha_1\alpha_2\alpha_3(3)}(x,p_\st^2) &=& -\frac{1}{2}\int\frac{d\varphi}{2\pi}\ \frac{p_\st^{\alpha_1\alpha_2\alpha_3}}{M^3}\left\lgroup \Gamma^{[+,+^\dagger]}(x,p_\st)-\Gamma^{[-,-^\dagger]}(x,p_\st)\right\rgroup + \widetilde\Gamma_{\{\partial\partial G\},1}^{\alpha_1\alpha_2\alpha_3(3)}(x,p_\st^2), \label{e:cor-GGG1} \\
\Gamma_{GGG,2}^{\alpha_1\alpha_2\alpha_3(3)}(x,p_\st^2) &=& -\frac{1}{2}\int\frac{d\varphi}{2\pi}\ \frac{p_\st^{\alpha_1\alpha_2\alpha_3}}{M^3}\left\lgroup \Gamma^{[+,-^\dagger]}(x,p_\st)-\Gamma^{[-,+^\dagger]}(x,p_\st)\right\rgroup + \widetilde\Gamma_{\{\partial\partial G\},2}^{\alpha_1\alpha_2\alpha_3(3)}(x,p_\st^2), \label{e:cor-GGG2} \\
\Gamma_{GGG,3}^{\alpha_1\alpha_2\alpha_3(3)}(x,p_\st^2) &=& -\frac{1}{12}\int\frac{d\varphi}{2\pi}\ \frac{p_\st^{\alpha_1\alpha_2\alpha_3}}{M^3}\left\lgroup \Gamma^{[+,+^\dagger(\Box)(\Box^\dagger)]}(x,p_\st)-\Gamma^{[-,-^\dagger(\Box)(\Box^\dagger)]}(x,p_\st)\right\rgroup \nonumber \\
&&+\frac{1}{6}\,\widetilde\Gamma_{\{\partial\partial G\},1}^{\alpha_1\alpha_2\alpha_3(3)}(x,p_\st^2)+\frac{1}{6}\,\Gamma_{GGG,1}^{\alpha_1\alpha_2\alpha_3(3)}(x,p_\st^2), \label{e:cor-GGG3} \\
\Gamma_{GGG,4}^{\alpha_1\alpha_2\alpha_3(3)}(x,p_\st^2) &=& -\frac{1}{12}\int\frac{d\varphi}{2\pi}\ \frac{p_\st^{\alpha_1\alpha_2\alpha_3}}{M^3}\left\lgroup \Gamma^{[+,-^\dagger(\Box)(\Box^\dagger)]}(x,p_\st)-\Gamma^{[-,+^\dagger(\Box)(\Box^\dagger)]}(x,p_\st)\right\rgroup \nonumber \\
&&+\frac{1}{6}\,\widetilde\Gamma_{\{\partial\partial G\},2}^{\alpha_1\alpha_2\alpha_3(3)}(x,p_\st^2)+\frac{1}{6}\,\Gamma_{GGG,2}^{\alpha_1\alpha_2\alpha_3(3)}(x,p_\st^2), \label{e:cor-GGG4} \\
\Gamma_{GGG,5}^{\alpha_1\alpha_2\alpha_3(3)}(x,p_\st^2) &=& -\frac{1}{2}\int\frac{d\varphi}{2\pi}\ \frac{p_\st^{\alpha_1\alpha_2\alpha_3}}{M^3}\left\lgroup \Gamma^{[+,+^\dagger(\Box)]}(x,p_\st)-\Gamma^{[-,-^\dagger(\Box^\dagger)]}(x,p_\st)\right\rgroup \nonumber \\
&&+\widetilde\Gamma_{\{\partial\partial G\},1}^{\alpha_1\alpha_2\alpha_3(3)}(x,p_\st^2) +\Gamma_{GGG,1}^{\alpha_1\alpha_2\alpha_3(3)}(x,p_\st^2) +3\,\Gamma_{GGG,3}^{\alpha_1\alpha_2\alpha_3(3)}(x,p_\st^2), \label{e:cor-GGG5} \\
\Gamma_{GGG,6}^{\alpha_1\alpha_2\alpha_3(3)}(x,p_\st^2)&-&\Gamma_{GGG,7}^{\alpha_1\alpha_2\alpha_3(3)}(x,p_\st^2) \nonumber \\
&=&-\frac{1}{6}\int\frac{d\varphi}{2\pi}\ \frac{p_\st^{\alpha_1\alpha_2\alpha_3}}{M^3}\left\lgroup \Gamma^{[(F(\xi)\Box),(F(0)\Box^\dagger)]}(x,p_\st)-\Gamma^{[(F(\xi)\Box^\dagger),(F(0)\Box)]}(x,p_\st)\right\rgroup \label{e:cor-GGG67}.
\eea
As can be seen in Eq.~\ref{e:cor-GGG67}, the correlators $\Gamma_{GGG,6}^{\{\alpha_1\alpha_2\alpha_3\}}(x,p_\st^2)$ and $\Gamma_{GGG,7}^{\{\alpha_1\alpha_2\alpha_3\}}(x,p_\st^2)$ appear as a specific combination only. The equations for the T-odd rank 3 pieces can be simplified further by using that the expressions for the T-even parts are zero. We nevertheless give the full results, since it will be a nice test for lattice calculations to verify whether these relations indeed are zero.

\end{document}